RU97-10-B
%
%
\headline{\hfil \folio}
\hoffset=0.5truein
\hsize=5.5truein
\vsize=8truein
%
\catcode`@=11                           
\newskip\ttglue
\def\ninefonts{%
   \global\font\ninerm=cmr9%
   \global\font\ninei=cmmi9%
   \global\font\ninesy=cmsy9%
   \global\font\nineex=cmex10%
   \global\font\ninebf=cmbx9%
   \global\font\ninesl=cmsl9%
   \global\font\ninett=cmtt9%
   \global\font\nineit=cmti9%
   \skewchar\ninei='177%
   \skewchar\ninesy='60%
   \hyphenchar\ninett=-1%
   \moreninefonts
   \gdef\ninefonts{\relax}}%
\def\moreninefonts{\relax}                      


\def\elevenfonts{%
   \global\font\elevenrm=cmr10 scaled \magstephalf%
   \global\font\eleveni=cmmi10 scaled \magstephalf%
   \global\font\elevensy=cmsy10 scaled \magstephalf%
   \global\font\elevenex=cmex10%
   \global\font\elevenbf=cmbx10 scaled \magstephalf%
   \global\font\elevensl=cmsl10 scaled \magstephalf%
   \global\font\eleventt=cmtt10 scaled \magstephalf%
   \global\font\elevenit=cmti10 scaled \magstephalf%
   \global\font\elevenss=cmss10 scaled \magstephalf%
   \skewchar\eleveni='177%
   \skewchar\elevensy='60%
   \hyphenchar\eleventt=-1%
   \moreelevenfonts
   \gdef\elevenfonts{\relax}}%
\def\moreelevenfonts{\relax}

\def\twelvefonts{%
   \global\font\twelverm=cmr10 scaled \magstep1%
   \global\font\twelvei=cmmi10 scaled \magstep1%
   \global\font\twelvesy=cmsy10 scaled \magstep1%
   \global\font\twelveex=cmex10 scaled \magstep1%
   \global\font\twelvebf=cmbx10 scaled \magstep1%
   \global\font\twelvesl=cmsl10 scaled \magstep1%
   \global\font\twelvett=cmtt10 scaled \magstep1%
   \global\font\twelveit=cmti10 scaled \magstep1%
   \global\font\twelvess=cmss10 scaled \magstep1%
   \skewchar\twelvei='177%
   \skewchar\twelvesy='60%
   \hyphenchar\twelvett=-1%
   \moretwelvefonts
   \gdef\twelvefonts{\relax}}%
\def\moretwelvefonts{\relax}                    

\def\fourteenfonts{%
   \global\font\fourteenrm=cmr10 scaled \magstep2%
   \global\font\fourteeni=cmmi10 scaled \magstep2%
   \global\font\fourteensy=cmsy10 scaled \magstep2%
   \global\font\fourteenex=cmex10 scaled \magstep2%
   \global\font\fourteenbf=cmbx10 scaled \magstep2%
   \global\font\fourteensl=cmsl10 scaled \magstep2%
   \global\font\fourteenit=cmti10 scaled \magstep2%
   \global\font\fourteenss=cmss10 scaled \magstep2%
   \skewchar\fourteeni='177%
   \skewchar\fourteensy='60%
   \morefourteenfonts
   \gdef\fourteenfonts{\relax}}%
\def\morefourteenfonts{\relax}                  


\def\tenmibfonts{
   \global\font\tenmib=cmmib10%
   \global\font\tenbsy=cmbsy10%
   \skewchar\tenmib='177%
   \skewchar\tenbsy='60%
   \gdef\tenmibfonts{\relax}}

\def\elevenmibfonts{
   \global\font\elevenmib=cmmib10 scaled \magstephalf%
   \global\font\elevenbsy=cmbsy10 scaled \magstephalf%
   \skewchar\elevenmib='177%
   \skewchar\elevenbsy='60%
   \gdef\elevenmibfonts{\relax}}

\def\twelvemibfonts{
   \global\font\twelvemib=cmmib10 scaled \magstep1%
   \global\font\twelvebsy=cmbsy10 scaled \magstep1%
   \skewchar\twelvemib='177%
   \skewchar\twelvebsy='60%
   \gdef\twelvemibfonts{\relax}}

\def\fourteenmibfonts{
   \global\font\fourteenmib=cmmib10 scaled \magstep2%
   \global\font\fourteenbsy=cmbsy10 scaled \magstep2%
   \skewchar\fourteenmib='177%
   \skewchar\fourteenbsy='60%
   \gdef\fourteenmibfonts{\relax}}

\def\mib{
   \tenmibfonts%
   \textfont0=\tenbf\scriptfont0=\sevenbf%
   \scriptscriptfont0=\fivebf%
   \textfont1=\tenmib\scriptfont1=\seveni%
   \scriptscriptfont1=\fivei%
   \textfont2=\tenbsy\scriptfont2=\sevensy%
   \scriptscriptfont2=\fivesy}%

\def\ninepoint{\ninefonts               
   \def\rm{\fam0\ninerm}%
   \textfont0=\ninerm\scriptfont0=\sevenrm\scriptscriptfont0=\fiverm
   \textfont1=\ninei\scriptfont1=\seveni\scriptscriptfont1=\fivei
   \textfont2=\ninesy\scriptfont2=\sevensy\scriptscriptfont2=\fivesy
   \textfont3=\nineex\scriptfont3=\nineex\scriptscriptfont3=\nineex
   \textfont\itfam=\nineit\def\it{\fam\itfam\nineit}%
   \textfont\slfam=\ninesl\def\sl{\fam\slfam\ninesl}%
   \textfont\ttfam=\ninett\def\tt{\fam\ttfam\ninett}%
   \textfont\bffam=\ninebf
   \scriptfont\bffam=\sevenbf
   \scriptscriptfont\bffam=\fivebf\def\bf{\fam\bffam\ninebf}%
   \def\mib{\relax}%
   \tt\ttglue=.5emplus.25emminus.15em
   \normalbaselineskip=11pt
   \setbox\strutbox=\hbox{\vrule height 8pt depth 3pt width 0pt}%
   \normalbaselines\rm\singlespaced}%

\def\tenpoint{
   \def\rm{\fam0\tenrm}%
   \textfont0=\tenrm\scriptfont0=\sevenrm\scriptscriptfont0=\fiverm
   \textfont1=\teni\scriptfont1=\seveni\scriptscriptfont1=\fivei
   \textfont2=\tensy\scriptfont2=\sevensy\scriptscriptfont2=\fivesy
   \textfont3=\tenex\scriptfont3=\tenex\scriptscriptfont3=\tenex
   \textfont\itfam=\tenit\def\it{\fam\itfam\tenit}%
   \textfont\slfam=\tensl\def\sl{\fam\slfam\tensl}%
   \textfont\ttfam=\tentt\def\tt{\fam\ttfam\tentt}%
   \textfont\bffam=\tenbf
   \scriptfont\bffam=\sevenbf
   \scriptscriptfont\bffam=\fivebf\def\bf{\fam\bffam\tenbf}%
   \def\mib{%
      \tenmibfonts%
      \textfont0=\tenbf\scriptfont0=\sevenbf%
      \scriptscriptfont0=\fivebf%
      \textfont1=\tenmib\scriptfont1=\seveni%
      \scriptscriptfont1=\fivei%
      \textfont2=\tenbsy\scriptfont2=\sevensy%
      \scriptscriptfont2=\fivesy}%
   \tt\ttglue=.5emplus.25emminus.15em
   \normalbaselineskip=12pt
   \setbox\strutbox=\hbox{\vrule height 8.5pt depth 3.5pt width 0pt}%
   \normalbaselines\rm\singlespaced}%

\def\elevenpoint{\elevenfonts           
   \def\rm{\fam0\elevenrm}%
   \textfont0=\elevenrm\scriptfont0=\sevenrm\scriptscriptfont0=\fiverm
   \textfont1=\eleveni\scriptfont1=\seveni\scriptscriptfont1=\fivei
   \textfont2=\elevensy\scriptfont2=\sevensy\scriptscriptfont2=\fivesy
   \textfont3=\elevenex\scriptfont3=\elevenex\scriptscriptfont3=\elevenex
   \textfont\itfam=\elevenit\def\it{\fam\itfam\elevenit}%
   \textfont\slfam=\elevensl\def\sl{\fam\slfam\elevensl}%
   \textfont\ttfam=\eleventt\def\tt{\fam\ttfam\eleventt}%
   \textfont\bffam=\elevenbf
   \scriptfont\bffam=\sevenbf
   \scriptscriptfont\bffam=\fivebf\def\bf{\fam\bffam\elevenbf}%
   \def\mib{%
      \elevenmibfonts%
      \textfont0=\elevenbf\scriptfont0=\sevenbf%
      \scriptscriptfont0=\fivebf%
      \textfont1=\elevenmib\scriptfont1=\seveni%
      \scriptscriptfont1=\fivei%
      \textfont2=\elevenbsy\scriptfont2=\sevensy%
      \scriptscriptfont2=\fivesy}%
   \tt\ttglue=.5emplus.25emminus.15em
   \normalbaselineskip=13pt
   \setbox\strutbox=\hbox{\vrule height 9pt depth 4pt width 0pt}%
   \normalbaselines\rm\singlespaced}%

\def\twelvepoint{\twelvefonts\ninefonts 
   \def\rm{\fam0\twelverm}%
   \textfont0=\twelverm\scriptfont0=\ninerm\scriptscriptfont0=\sevenrm
   \textfont1=\twelvei\scriptfont1=\ninei\scriptscriptfont1=\seveni
   \textfont2=\twelvesy\scriptfont2=\ninesy\scriptscriptfont2=\sevensy
   \textfont3=\twelveex\scriptfont3=\twelveex\scriptscriptfont3=\twelveex
   \textfont\itfam=\twelveit\def\it{\fam\itfam\twelveit}%
   \textfont\slfam=\twelvesl\def\sl{\fam\slfam\twelvesl}%
   \textfont\ttfam=\twelvett\def\tt{\fam\ttfam\twelvett}%
   \textfont\bffam=\twelvebf
   \scriptfont\bffam=\ninebf
   \scriptscriptfont\bffam=\sevenbf\def\bf{\fam\bffam\twelvebf}%
   \def\mib{%
      \twelvemibfonts\tenmibfonts%
      \textfont0=\twelvebf\scriptfont0=\ninebf%
      \scriptscriptfont0=\sevenbf%
      \textfont1=\twelvemib\scriptfont1=\ninei%
      \scriptscriptfont1=\seveni%
      \textfont2=\twelvebsy\scriptfont2=\ninesy%
      \scriptscriptfont2=\sevensy}%
   \tt\ttglue=.5emplus.25emminus.15em
   \normalbaselineskip=14pt
   \setbox\strutbox=\hbox{\vrule height 10pt depth 4pt width 0pt}%
   \normalbaselines\rm\singlespaced}%

\def\fourteenpoint{\fourteenfonts\twelvefonts 
   \def\rm{\fam0\fourteenrm}%
   \textfont0=\fourteenrm\scriptfont0=\twelverm\scriptscriptfont0=\tenrm
   \textfont1=\fourteeni\scriptfont1=\twelvei\scriptscriptfont1=\teni
   \textfont2=\fourteensy\scriptfont2=\twelvesy\scriptscriptfont2=\tensy
   \textfont3=\fourteenex\scriptfont3=\fourteenex
      \scriptscriptfont3=\fourteenex
   \textfont\itfam=\fourteenit\def\it{\fam\itfam\fourteenit}%
   \textfont\slfam=\fourteensl\def\sl{\fam\slfam\fourteensl}%
   \textfont\bffam=\fourteenbf
   \scriptfont\bffam=\twelvebf
   \scriptscriptfont\bffam=\tenbf\def\bf{\fam\bffam\fourteenbf}%
   \def\mib{%
      \fourteenmibfonts\twelvemibfonts\tenmibfonts%
      \textfont0=\fourteenbf\scriptfont0=\twelvebf%
      \scriptscriptfont0=\tenbf%
      \textfont1=\fourteenmib\scriptfont1=\twelvemib%
      \scriptscriptfont1=\tenmib%
      \textfont2=\fourteenbsy\scriptfont2=\tenbsy%
      \scriptscriptfont2=\tenbsy}%
   \normalbaselineskip=17pt
   \setbox\strutbox=\hbox{\vrule height 12pt depth 5pt width 0pt}%
   \normalbaselines\rm\singlespaced}%
%
%

\def\singlespaced{
   \baselineskip=\normalbaselineskip}           


%
%
\twelvepoint
%
%
\def\begintitle{\begingroup%
\obeylines\fourteenpoint\bf\parindent=0.29truein}
\def\endtitle{\vglue 1truecm\endgroup}

\def\showheadline#1#2{\headline={\ifnum\pageno>1{\ifodd\pageno{\hfil\tenpoint #1\hfil} %
\else{\hfil\tenpoint #2\hfil}\fi} \else{\hfil}\fi}}

\def\beginauthor{\begingroup%
\obeylines\fourteenpoint\parindent=0.29truein}
\def\endauthor{\endgroup}

\def\address#1{\hbox to \hsize{\hglue 0.29in\relax
\vbox{\hsize=4.70in\relax\rightskip=0pt plus 1in\relax\noindent#1}\hfil}}

\long\def\beginaddress#1\endaddress{\vglue 6pt\address{#1}\vglue 24pt}

\def\beginabstract{\begingroup\leftskip=0.29in%
\tenpoint\noindent{\bf Abstract\ \ \ }}
\def\endabstract{\vskip 1pt minus1pt\endgroup}

\def\finalversion{\headline{\hfil}}

\def\section#1{\vskip 24pt plus4pt minus4pt\goodbreak\leftline{\bf #1}%
\vglue 12pt\nobreak\noindent\kern -0.0em}

\def\subsection#1{\vskip 12pt plus4pt minus4pt\goodbreak\leftline{\bf #1}%
\nobreak\noindent\kern -0.0em}

\def\subsubsection#1{\vskip 12pt plus4pt minus4pt\goodbreak\leftline{\it #1}%
\nobreak\noindent\kern -0.0em}

\def\begincaption#1{\begingroup\tenpoint\noindent#1\ \ \ }
\def\endcaption{\endgroup}

\newbox\@capbox                                 
\newcount\@caplines                             

\def\references{\section{REFERENCES}\tenpoint\parindent=0pt
\raggedright\rightskip=0pt plus 5em}

\def\ref#1#2{\hbox to \hsize{\vbox{\tenpoint\hsize=0.2in\relax #1\hfil}
\hfil\vtop{\hsize=4.75in\relax\tenpoint #2}}}
%
%
\vglue 1.0truein

\finalversion
%
%
\input epsf
%
%
%
%
\begintitle
Physical Properties of the Double Exchange
and Its Relevance to the Specific Heat of 
Double Perovskite Materials
\endtitle

\beginauthor
H. C. Ren
\endauthor

\beginaddress
Department of Physics, The Rockefeller University,
New York, NY 10021, USA
\endaddress

\beginauthor
M. K. Wu
\endauthor

\beginaddress
Department of Physics, National Tsing Hua University,
Hsinchu, Taiwan, ROC
\endaddress

\beginabstract

In response to the experimental discovery of double perovskite Ru-
base material, The magnetic and electronic properties of the double 
exchange mechanism are studied theoretically. 
The electron band structure and the spin wave spectrum
response at zero temperature with the canting background are computed 
from the first principle Hamiltonian. The long range Coulomb interaction 
between charge carriers is found to be crucial in stabilizing the canting 
order. A Monte Carlo simulation is performed on de-Gennes's effective 
magnetic Hamiltonian with classical spin approximation. The peaks of the 
specific heat as a function of temperature, which correspond to FM(AF) and 
canting transitions, and its response to the external magnetic field 
are consistent with the observation. For certain region of parameters, 
a more complex canting order than layer type is found at 
sufficiently low temperature. The effective interaction between carriers 
via exchanging a magnon wave is found to be rather similar to that via 
exchanging an acoustic phonon. The pros and cons to a genuine superconducting 
order in this system is analyzed. 

\endabstract

\section{1. Introduction}

Recently, a new type of materials of double perovskite structure is 
synthesized by [1], which displays a number of interesting 
physics [1-2]. The molecular formula are $Sr_2YRu_{1-x}
Cu_xO_6$, $Ba_2YRu_{1-x}Cu_xO_6$, etc.. The magnetic properties are 
unusual with two phase transitions as a function of temperature. 
The underlying mechanism was conjectured to be the double exchange 
effect [3-5], the same mechanism partially responsible to the colossal 
magnetoresistance in the single perovskite compound, $La_{2-x}Ca_xMnO_4$
[6]. In addition, the materials become superconducting for $x\ge 0.04$ 
and its transition temperature reaches $T_C =20-30K$ for $x=0.1$.
The coherence length is about $\xi=35\AA$ and the super phase possess 
several novel features different from cuprates. 

The double perovskite crystal structure of the parent compound 
$(Sr, Ba)_2YRuO_6$ can be approximated by a cubic structure with the 
side length of a fundamental cube $l\sim 8.16\AA$. The $(Ru, Cu)$ ions 
are located at the face-center sites, the $Y$ ions are sitting at the 
octahedral sites and $Sr$ ions at tetrahedral sites. The $O$ ions are 
placed at the middle of the shortest $Y-Ru$ bond. A small fraction of 
$Ru$ is replaced by $Cu$ through doping.

The charge balance of the ions $Sr(Ba)$, $Y$, $Ru$ and $O$ of the parent 
compound are $+2$, $+3$, $+5$ and $-2$. Except $Ru^{+5}$, the outer 
electronic configurations of all other ions are closed shells of the 
continuous rotation group. The configuration of $Ru^{+5}$ is $4d^3$. It is 
likely that the strong Hund's rule coupling enforces the spins of the three 
$d$ electrons parallel with total spin $S=3/2$ and their orbital wave 
function fill up a multiplet of the $T_2$ representation of the cubic 
group, supported by $xy$, $yz$ and $zx$ (notably, the lobes of these
wave functions point to the directions of the twelve nearest neighbors 
of a face center site). With only the nearest
neighbor antiferromagnetic exchange, the Neel state may assume a 
layer structure [7] perpendicular to one of the 4-fold axis with the 
spins of $Ru^{+5}$ alternating along successive layers. This is consistent 
with the tetragonal symmetry observed by X-ray diffraction data [1]. 
Replacing part of $Ru$ ions offset the charge 
balance and creates itinerant holes with density $2x$ per molecule of 
$(Sr, Ba)_2YRu_{1-x}Cu_xO_6$. This may also break the symmetry further 
to an orthohmobic structure as indicated by the X-ray data of the doped 
compounds.

With itinerant holes, the state of a $Ru$ ion is a superposition of 
states with charge +5, as in the parent compound and states of charge 
+6. In the latter case, Hund's implies a total spin $S=1$. Regarding 
the latter states as combinations of the product states of an ion core 
$Ru^{+5}$ of spin $S={3\over 2}$ and a hole of spin ${1\over 2}$. 
Houd's rule will prefer antiparallel spins of the ion core and the 
hole. The Hamiltonian combining both the antiferromagnetic exchange 
and the double exchange reads
$$H=\sum_{ij}J_{ij}\vec S_i\cdot\vec S_j-{1\over 2}\sum_{ij}
t_{ij}(\Psi_i^\dagger \Psi_j+\Psi_j^\dagger \Psi_i)
+{1\over 2}\lambda\sum_j\vec S_j \Psi_j^\dagger\vec\sigma \Psi_j
+U_{Coul.},\eqno(1.1)$$
where $\vec S_j$ is the total spin operator of the ion core, $\Psi_j$,
$\Psi_j^\dagger$ are the creation and the annihilation operators of 
the itinerant holes, each of two components with spin up and spin 
down, $J_{ij}$ denotes the antiferromagnetic exchange, $t_{ij}$ denotes
the hopping amplitude of the itinerant electrons and $\lambda$ denotes 
the exchange interaction of the itinerant electrons with the ion cores.
The Coulomb interaction, $U_{Coul.}$, of (1.1) is given by 
$$U_{Coul.}={e^2\over 8\pi\varepsilon}\Big(U\sum_{j,\sigma,\sigma^\prime}
:\Psi_{j\sigma}^\dagger\Psi_{j\sigma}
\Psi_{j\sigma^\prime}^\dagger\Psi_{j\sigma^\prime}:+
\sum_{i\neq j,\sigma,\sigma^\prime} {:\Psi_{i\sigma}^\dagger\Psi_{i\sigma}
\Psi_{j\sigma^\prime}^\dagger\Psi_{j\sigma^\prime}:\over 
|\vec R_i-\vec R_j|}\Big),\eqno(1.2)$$  where the first term represents the 
on-site Coulomb repulsion, $\varepsilon$ denotes the dielectric 
constant of the ion cores and :(...): enforce the normal ordering. 
As we shall see, the long range Coulomb 
interaction is rather important to maintain a homogeneous canting magnetic 
order against clustering. The magnetic anisotropicity, 
the possible Jahn-Teller effect due to the degeneracy of $d$-band, 
as well as the impurity effect, which might be relevant in real systems, 
have been neglected for simplicity. 

It was pointed out by Anderson and Hasegawa [4] that the double exchange 
energy, steming from intra-atomic Hund's rule coupling, ought to be much 
stronger than the hopping energy, i.e., $\lambda S>>t$. In these 
circumstance, the effective magnetic Hamiltonian after integrating 
out the electronic degrees of freedom can be approximated by [6]
$$H_{\rm{eff.}}=\sum_{ij}J_{ij}\vec S_i\cdot\vec S_j-
\sum_{ij}b_{ij}\cos{\Theta_{ij}\over 2}\eqno(1.3)$$
with $$\cos{\Theta_{ij}\over 2}=\sqrt{{1\over 2}+{\vec S_i\cdot
\vec S_j\over 2S^2}}\eqno(1.4)$$ for low carrier densities

This paper addresses mainly the theoretical issues of the magnetic 
properties associated with the double exchange effect. 
Throughout this paper, we shall adopt tetragonal order for the 
magnetic states, which preserves the maximum subgroup symmetry 
of the original cubic group. Only $J_{ij}$'s and $t_{ij}$'s for 
nearest neighboring bond are kept different from zero for simplicity
and we introduce the notations that
$$J_{ij}=\cases{J,& for interlayer bond $<ij>$; \cr J^\prime,
& for intralayer bond $<ij>$\cr}\eqno(1.5)$$ and  
$$t_{ij}=\cases{t & for interlayer bond $<ij>$; \cr t^\prime,
& for intralayer bond $<ij>$\cr}\eqno(1.6)$$
Accordingly, the parameters $b_{ij}$ for the effective magnetic 
Hamiltonian (1.3) read $$b_{ij}=\cases{b & for interlayer bond $<ij>$; 
\cr b^\prime, & for intralayer bond $<ij>$\cr}.\eqno(1.7)$$ 
On a cubic lattice of ${\cal N}$ sites with a periodic boundary 
condition, there are $2{\cal N}$ intra-layer bonds and $4{\cal N}$
interlayer bond. Following de Gennes 
[5], we a homogeneous spin orientation in each layer and a canting 
order between successive layers. The canting angle as follows the 
minimization of (1.3) is $$\cos{\Theta\over 2}={b\over 4JS^2}.\eqno(1.8)$$

This paper is organized as follows. In the section 2, we shall expand 
the Hamiltonian (1.1) around a canting order and examine the band 
structure of the itinerant holes. The first principle calculation of 
the magnon spectrum at zero temperature will be displayed in the section 3. 
The Monte Carlo simulation of the magnetic Hamiltonian (1.3) at 
arbitrary temperatures is presented in the section 4. The section 
5 is devoted to a sketch of possible coupling between the magnon 
and a superconducting order. In the section 5, we shall come back to the 
real systems and discuss the possibility of a genuine superconducting 
order in them.

\section{2. Mean-Field Expansion of the Hamiltonian around 
Canting}

Denote the side length of a basic cube by $l$ and the cubic axes, 
(100), (010) and (001) by $\hat x,\hat y, \hat z$. The locations 
magnetic ions are given by 
$$\vec R=n_1\vec e_1+n_2\vec e_2+n_3\vec e_3,
\eqno(2.1)$$ where $n_1$, $n_2$ and $n_3$ are integers,
$$\vec e_1={l\over 2}(\hat y+\hat z),\eqno(2.2)$$ 
$$\vec e_2={l\over 2}(\hat z+\hat x),\eqno(2.3)$$ 
$$\vec e_3={l\over 2}(\hat x+\hat y),\eqno(2.4)$$ 
and the total number of sites is ${\cal N}$. Adapting the 
tetragonal layer magnetic ordering, we divide the lattice 
into two sublattices shown in Fig 1. with $n_1+n_2={\rm{even}}$ on the 
sublattice $A$ and $n_1+n_2={\rm{odd}}$ on the sublattice $B$. 
The expectation value of the ion spins of the sublattice $A$ is 
denoted by $\vec S_A=S\vec\zeta_A$ and that of the sublattice $B$ 
by $\vec S_B=S\vec\zeta_B$. We choose the $\zeta$-axis along the 
direction of $\vec S_A+\vec S_B$ and the $\eta$-axis as 
$$\hat\eta={\vec \zeta_A\times\vec \zeta_B\over\sin\Theta}
\eqno(2.5)$$ with $\Theta$ the mutual angle between $\vec S_A$ and 
$\vec S_B$. We introduce further
$$\hat\xi=\hat\eta\times\hat\zeta,\eqno(2.6)$$
$$\hat\xi_A=\hat\eta\times\hat\zeta_A\eqno(2.7)$$
and $$\hat\xi_B=\hat\eta\times\hat\zeta_B.\eqno(2.8)$$
The sets ($\hat\xi, \hat\eta, \hat\zeta$), ($\hat\xi_A, \hat\eta, \hat
\zeta_A$) and ($\hat\xi_B, \hat\eta, \hat\zeta_B$) each forms a right hand 
coordinate system. The first of them will be referred to as the spin frame. 
Since the magnetic anisotropic energy has been neglected, the orientation 
of the spin frames does not couple with the crystal axis. The relevance 
of the magnetic anisotropy will be discussed in the Appendix A. 

\topinsert
\hbox to\hsize{\hss
	\epsfxsize=4.0truein\epsffile{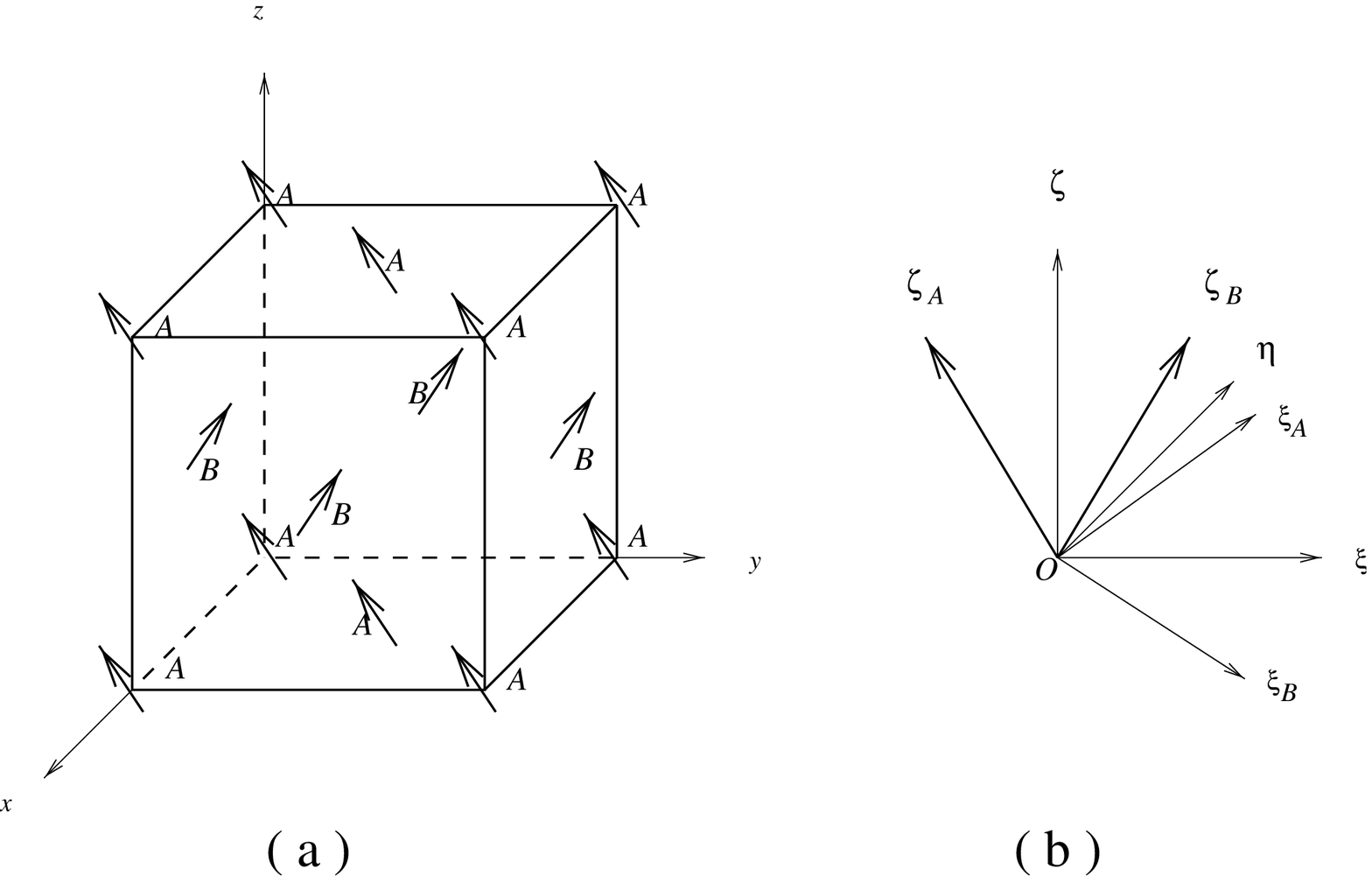}
	\hss}
\begincaption{Figure 1}
(a) The magnetic ordering. The magnetic sites of the sublattices $A$ and 
$B$ are marked explicitly. (b) The spin frame. The angle between $\zeta_A$,
and $\zeta_B$ axes is $\Theta$. 
\endcaption
\endinsert

With large spin approximation [8], we write 
$$\vec S_j=\sqrt{{S\over 2}}(c_j+c_j^\dagger)\hat\xi_A
-i\sqrt{{S\over 2}}(c_j-c_j^\dagger)\hat\eta
+(S-c_j^\dagger c_j)\hat\zeta_A+...\eqno(2.9)$$ for the site $j$ on the 
sublattice $A$ and  
$$\vec S_j=\sqrt{{S\over 2}}(d_j+d_j^\dagger)\hat\xi_B
-i\sqrt{{S\over 2}}(d_j-d_j^\dagger)\hat\eta
+(S-d_j^\dagger d_j)\hat\zeta_B+...\eqno(2.10)$$ for the site $j$ on the 
sublattice $B$, where $c_j$, $c_j^\dagger$, $d_j$ and $d_j^\dagger$
satisfy the commutation relations of elementary boson operators.
After some algebra, which is shown in detail in Appendix B, the 
Hamiltonian (1.1) is expanded as
$$H=H_0+H_1+H_2+H_3+H_4,\eqno(2.11)$$ where 
$$H_0=2{\cal N}S^2(J^\prime+2J\cos\Theta),\eqno(2.12)$$
$$H_1=4JS^{3\over 2}\sqrt{2{\cal N}}
(\zeta_0+\zeta_0^\dagger)\sin\Theta,\eqno(2.13)$$
$$H_2=\sum_{\vec p}\psi_{\vec p}^\dagger(E_{\vec p}^\prime-\mu)
\psi_{\vec p}+\sum_{\vec p}(\nu_{\vec p}^+\zeta_{\vec p}^\dagger \zeta_{\vec p}
+\nu_{\vec p}^-z_{\vec p}^\dagger z_{\vec p})$$
$$-JS\sin^2{\Theta\over 2}\sum_{\vec p}v_{\vec p}
(z_{\vec p}z_{-\vec p}+z_{-\vec p}^\dagger z_{\vec p}^\dagger
-\zeta_{\vec p}\zeta_{-\vec p}-\zeta_{-\vec p}^\dagger\zeta_{\vec p}^\dagger)
\eqno(2.14)$$ with $\nu_{\vec p}^\pm=2S\Big[J^\prime(-2+u_{\vec p})
-4J\cos\Theta\mp Jv_{\vec p}\cos^2{\Theta\over 2}\Big]$,
$$H_3=-\lambda\sqrt{{S\over 2{\cal N}}}\sum_{\vec p,\vec q}\Big[
\zeta_{\vec q}\psi_{\vec p+{\vec q\over 2}}^\dagger\tau_-
\psi_{\vec p-{\vec q\over 2}}+\zeta_{\vec q}^\dagger
\psi_{\vec p-{\vec q\over 2}}^\dagger\tau_+\psi_{\vec p+{\vec q\over 2}}$$
$$+z_{\vec q}\psi_{\vec p+{\vec q\over 2}}^\dagger\rho_1\tau_-
\psi_{\vec p-{\vec q\over 2}}+z_{\vec q}^\dagger
\psi_{\vec p-{\vec q\over 2}}^\dagger\rho_1\tau_+
\psi_{\vec p+{\vec q\over 2}}\Big]+\hbox{three-magnon terms},\eqno(2.15)$$
$$H_4=-{\lambda\over 2{\cal N}}\sum_{\vec p,\vec p^\prime,\vec q}
\Big[(\zeta_{\vec p^\prime-{\vec q\over 2}}^\dagger\zeta_{\vec p^\prime+
{\vec q\over 2}}+z_{\vec p^\prime-{\vec q\over 2}}^\dagger z_{\vec p^\prime+
{\vec q\over 2}})\psi_{\vec p+{\vec q\over 2}}^\dagger\tau_3
\psi_{\vec p-{\vec q\over 2}}$$ 
$$+(\zeta_{\vec p^\prime-{\vec q\over 2}}^\dagger 
z_{\vec p^\prime+{\vec q\over 2}}+z_{\vec p^\prime-{\vec q\over 2}}^\dagger 
\zeta_{\vec p^\prime+{\vec q\over 2}})\psi_{\vec p+{\vec q\over 2}}^\dagger
\rho_1\tau_3\psi_{\vec p-{\vec q\over 2}}\Big]+\hbox{four-magnon terms}
\eqno(2.16)$$ and $$U_{Coul.}={e^2\over {\cal N}\varepsilon}\sum_{\vec p,
\vec p^\prime,\vec q}\Big(D_{\vec q}^{(0)}\psi_{\vec p+{\vec q\over 2}}^
\dagger\psi_{\vec p^\prime-{\vec q\over 2}}^\dagger
\psi_{\vec p^\prime+{\vec q\over 2}}\psi_{\vec p-{\vec q\over 2}}$$
$$+D_{\vec q}^{(0)\prime}(\rho_1)_{\alpha\beta}(\rho_1)_{\alpha^\prime
\beta^\prime}\psi_{\vec p+{\vec q\over 2}\alpha}^\dagger
\psi_{\vec p^\prime-{\vec q\over 2}\alpha^\prime}^\dagger
\psi_{\vec p^\prime+{\vec q\over 2}\beta^\prime}
\psi_{\vec p-{\vec q\over 2}\beta}\Big).\eqno(2.17)$$ In (2.15)-(2.16), 
only the terms we shall use are exhibited explicitly. The magnon operators
$\zeta_{\vec p}$ and $z_{\vec p}$ of (2.13)-(2.16) are related to $c_j$ and 
$d_j$ via (B.1), (B.2), (B.19) and (B.20) in the appendix. The hole operator 
$\psi_{\vec p}$ of (2.14)-(2.17) is a $4\times 1$ column matrix and is related 
to the $2\times 1$ column matrix $\Psi_j$ 
through (B.3), (B.4) and (B.21). The $4\times 4$ Dirac matrices
$$\tau_1=\left(\matrix{\sigma_1&0\cr 0&\sigma_1}\right),
\tau_2=\left(\matrix{\sigma_2&0\cr 0&\sigma_2}\right),
\tau_3=\left(\matrix{\sigma_3&0\cr 0&\sigma_3}\right)\eqno(2.18)$$
and $$\rho_1=\left(\matrix{0&I\cr I&0}\right),
\rho_2=\left(\matrix{0&-iI\cr iI&0}\right), 
\rho_3=\left(\matrix{I&0\cr 0&-I}\right)\eqno(2.19)$$
with $$[\rho_a,\tau_b]=0\eqno(2.20)$$ and $$\tau_{\pm}={1\over 2}(\tau_1\pm
i\tau_2)\eqno(2.21)$$ have been introduced accordingly, 
where $I$ is a $2\times 2$ unit matrix and $\sigma_a$'s are Pauli matrices.
The Bloch momentum $\vec p(\vec p^\prime,\vec q)$ throughout (2.13-17)
corresponds to the translational invariance 
within each sublattice. In terms of the dimensionless papameters
$$\alpha_1={l\over 2}(p_x+p_y),\eqno(2.22)$$
$$\alpha_2={l\over 2}(-p_x+p_y)\eqno(2.23)$$ and $$\alpha_3=p_zl,\eqno(2.24)$$
the Brillouin zone is specified as $\pi<\alpha_1, \alpha_2, \alpha_3\leq\pi$,
the function $u_{\vec p}$ and $v_{\vec p}$ of (2.14) is given by
$$u_{\vec p}=\cos\alpha_1+\cos\alpha_2\eqno(2.25)$$ and 
$$v_{\vec p}=4\cos{\alpha_1\over 2}\cos{\alpha_2\over 2}
\cos{\alpha_3\over 2},\eqno(2.26)$$ and the bare Coulomb propagators 
$D_{\vec q}^{(0)}$ and $D_{\vec q}^{(0)\prime}$ of (2.17) read 
$$D_{\vec q}^{(0)}={4\over l(2\alpha_\perp^2+\alpha_3^2)}+U-{0.36485\over l}
\eqno(2.27)$$ and $$D_{\vec q}^{(0)\prime}=U-{0.17943\over l}.\eqno(2.28)$$

The Hamiltonian (2.11)-(2.17) is invariant under the tetragonal crystal 
rotation, inversion, and the ${\cal C}$-transformation introduced in the
appendix B, $${\cal C}z_{\vec p}{\cal C}^{-1}=-z_{\vec p}\eqno(2.29)$$ 
$${\cal C}\zeta_{\vec p}{\cal C}^{-1}=\zeta_{\vec p}\eqno(2.30)$$ 
and $${\cal C}\psi_{\vec p}{\cal C}^{-1}=-\rho_3\psi_{\vec p}.\eqno(2.31)$$

The matrix of the hole kinetic energy $E_{\vec p}$ is given by 
$$E_{\vec p}-\mu=-t^\prime u_{\vec p}-tv_{\vec p}\rho_3\Big(
\tau_3\cos{\Theta\over 2}-\tau_1\sin{\Theta\over 2}\Big)
+{1\over 2}\lambda S\tau_3-\mu,\eqno(2.32)$$
which is easily diagonalized and produces four energy bands:
$$\epsilon_{\vec p,s}^+=-t^\prime(u_{\vec p}+\mu)-s\Delta_{\vec p}^+
\eqno(2.33)$$ for ${\cal C}=1$ and $$\epsilon_{\vec p,s}^-=
-t^\prime(u_{\vec p}+\mu)-s\Delta_{\vec p}^-
\eqno(2.34)$$ for ${\cal C}=-1$, where $s=\pm 1$ and  
$$\Delta_{\vec p}^\pm=\sqrt{{1\over 4}\lambda^2S^2\pm\lambda 
Stv_{\vec p}\cos{\Theta\over 2}+t^2v_{\vec p}^2}.\eqno(2.35)$$
The wave functions corresponding to the four bands are
$${\cal U}_{\vec p,s}^+=\left(\matrix{\phi_{\vec p,s}^+\cr 0\cr}\right)
\eqno(2.36)$$ for $\epsilon_{\vec p,s}^+$ and 
$${\cal U}_{\vec p,s}^-=\left(\matrix{0\cr \phi_{\vec p,s}^-\cr}\right)
\eqno(2.37)$$ for $\epsilon_{\vec p,s}^-$, where $2\times 1$ matrices 
$\phi_{\vec p,s}^{\pm}$ are defined by
$${1\over \Delta_{\vec p}^\pm}\Big[\Big({1\over 2}\lambda S
\pm tv_{\vec p}\cos{\Theta\over 2}\Big)\sigma_3\mp tv_{\vec p}\sigma_1
\sin{\Theta\over 2}\Big)\phi_{\vec p,s}^{\pm}=s\phi_{\vec p,s}^{\pm}.
\eqno(2.38)$$ The electron spin ceases to be a good quantum number and the 
only degeneracies in Brillouin zone correspond the points related by the 
space inversion and the proper tetragonal rotations. 
The density of states of the four bands are plotted in Fig. 2. 

\topinsert
\hbox to\hsize{\hss
	\epsfxsize=3.0truein\epsffile{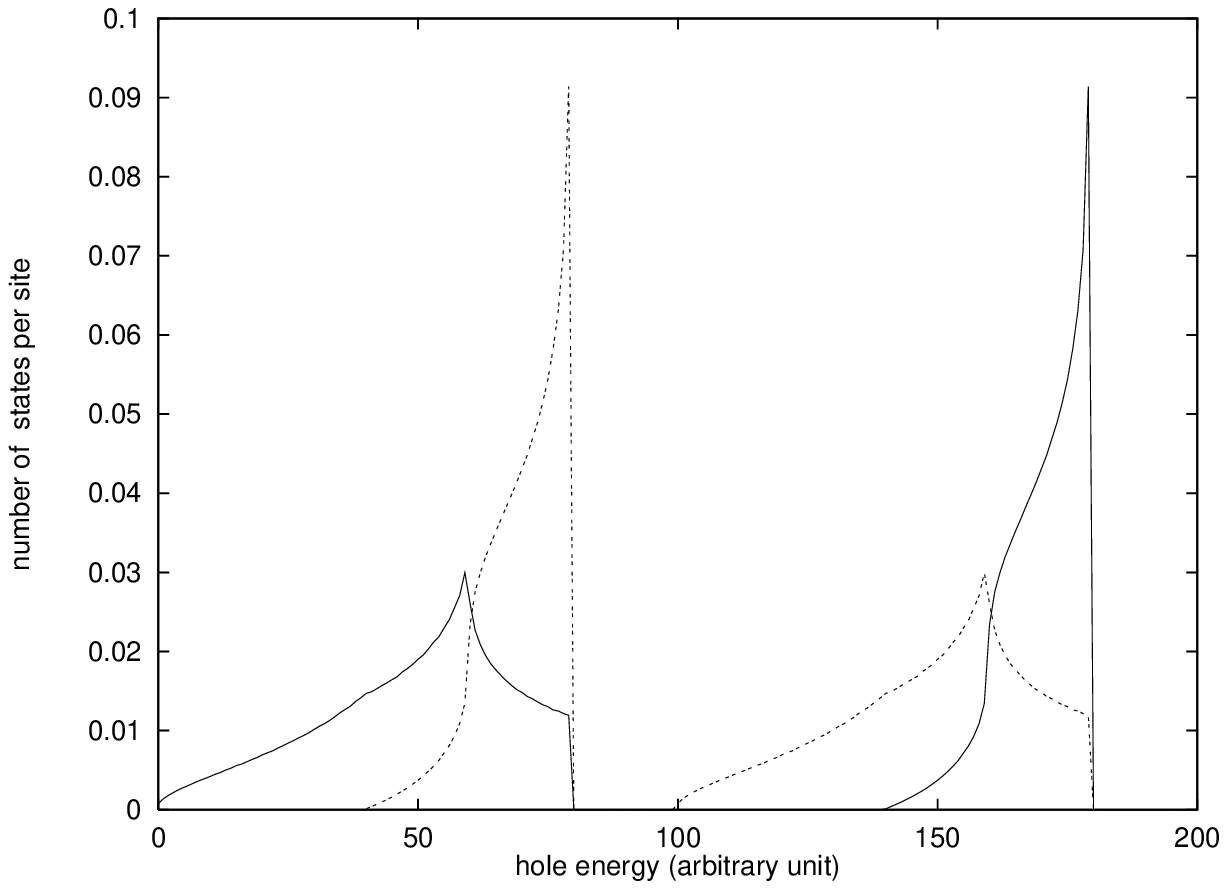}
	\hss}
\centerline{(a)}
\hbox to\hsize{\hss
	\epsfxsize=3.0truein\epsffile{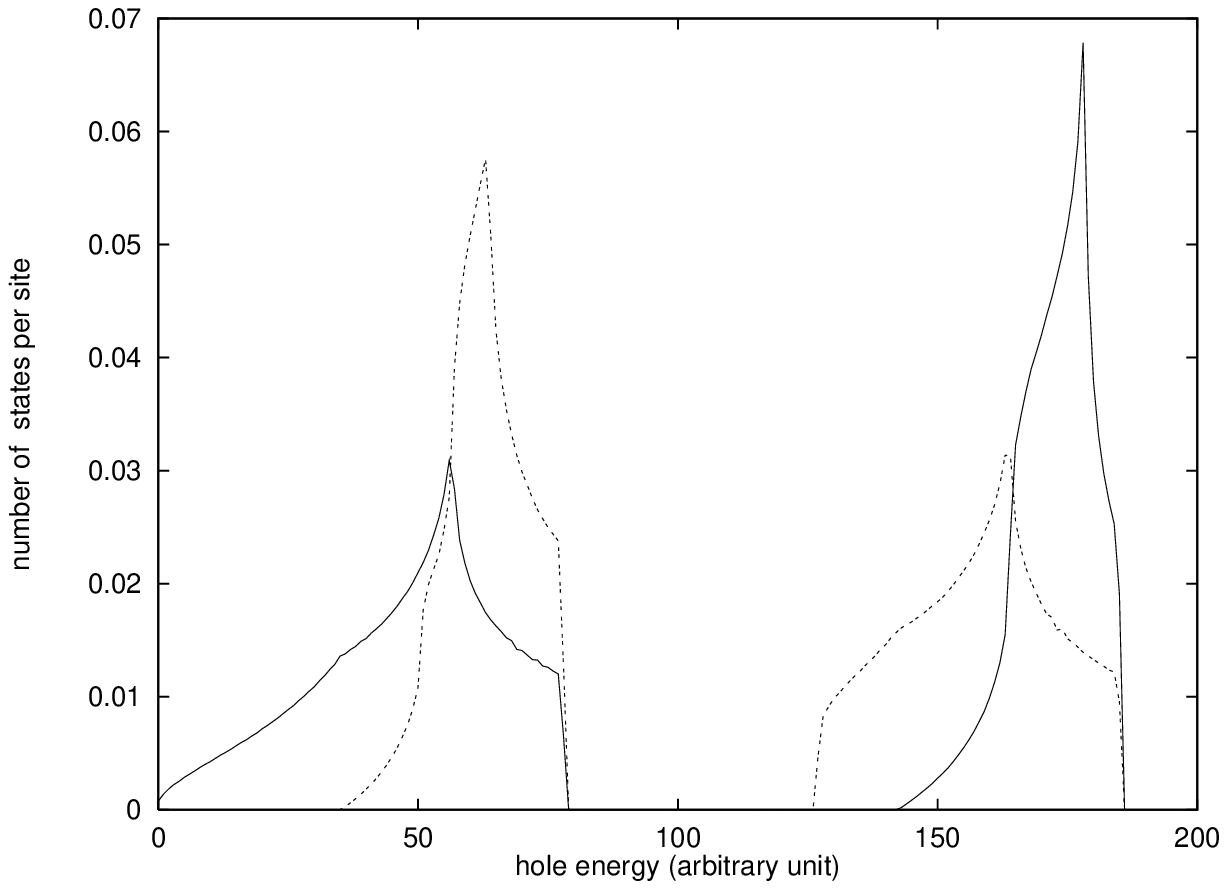}
	\hss}
\centerline{(b)}
\hbox to\hsize{\hss
	\epsfxsize=3.0truein\epsffile{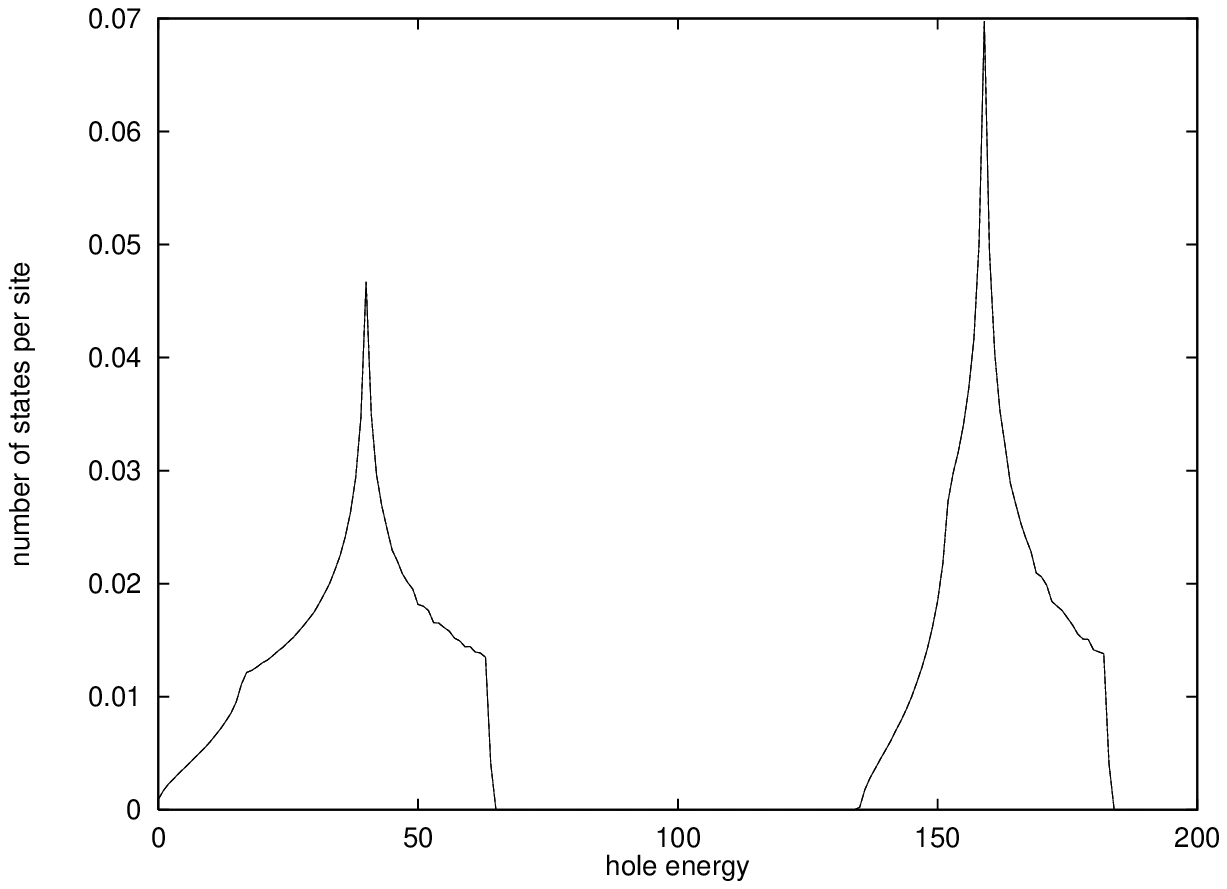}
	\hss}
\centerline{(c)}
\begincaption{Figure 2}
The density of states of itinerant holes at several canting angles. 
(a) $\Theta=0$; (b) $\Theta=90^\circ$; (c) $\Theta=180^\circ$. The solid 
line represents the band with ${\cal C}=1$. and the dashed line 
represents the band with ${\cal C}=-1$. Both kinds of bands degenerate 
for the pure antiferromagnetic order (c)
\endcaption
\endinsert

The free electron propagator with energy $p_0$ and momentum $\vec p$ is 
given by: $${\cal S}(p_0,\vec p)=\int_{-\infty}^\infty dt e^{ip_0t}
<|T\psi_{\vec p}(t)\psi_{\vec p}(0)^\dagger|>=S^+(p_0,\vec p)\rho_-
+S^-(p_0,\vec p)\rho_+
\eqno(2.39)$$ $$S^\pm(p_0,\vec p)=i{p_0+t^\prime u_{\vec p}\pm 
tv_{\vec p}\tau_3^\prime+{1\over 2}\lambda S\tau_3\over (p_0+t^\prime 
u_{\vec p})^2-\Delta_{\vec p}^{\pm 2}}\eqno(2.40)$$ with
$$\tau_3^\prime=\tau_3\cos{\Theta\over 2}-\tau_1\sin{\Theta\over 2}
\eqno(2.41)$$ and $T...$ the time ordering symbol. 
The state $|>$ in (2.39) denotes the Fermi sea of the itinerant holes
and the time evolution of $\psi_{\vec p}$ is generated by the quadratic
hole Hamiltonian.   

\section{3. Magnetic Energies with Normal Holes}

In this section we shall calculate the ground state energy and 
the spin wave spectrum at zero temperature using the Hamiltonian 
(2.11)-(2.17). For the sake of simplicity, we assume that only the band 
$\epsilon_{\vec p,+}^+$ is partially filled and denote
$$\epsilon_{\vec p,+}^+\equiv \epsilon_{\vec p}=-t^\prime u_{\vec p}
-\Delta_{\vec p}-\mu\eqno(3.1)$$ with 
$$\Delta_{\vec p}=\sqrt{{1\over 4}\lambda^2S^2+\lambda S tv_{\vec p}
\cos{\Theta\over 2}+t^2v_{\vec p}^2}.\eqno(3.2)$$ As we shall see in 
the next section, the incorporation of possible superconductivity 
orders does not modify the result drastically. We confine ourselves 
within the normal phase of the carriers in this section. 

The basic ingredients of the Feynman diagrams are displayed in Fig. 3 
\topinsert
\hbox to\hsize{\hss
	\epsfxsize=4.0truein\epsffile{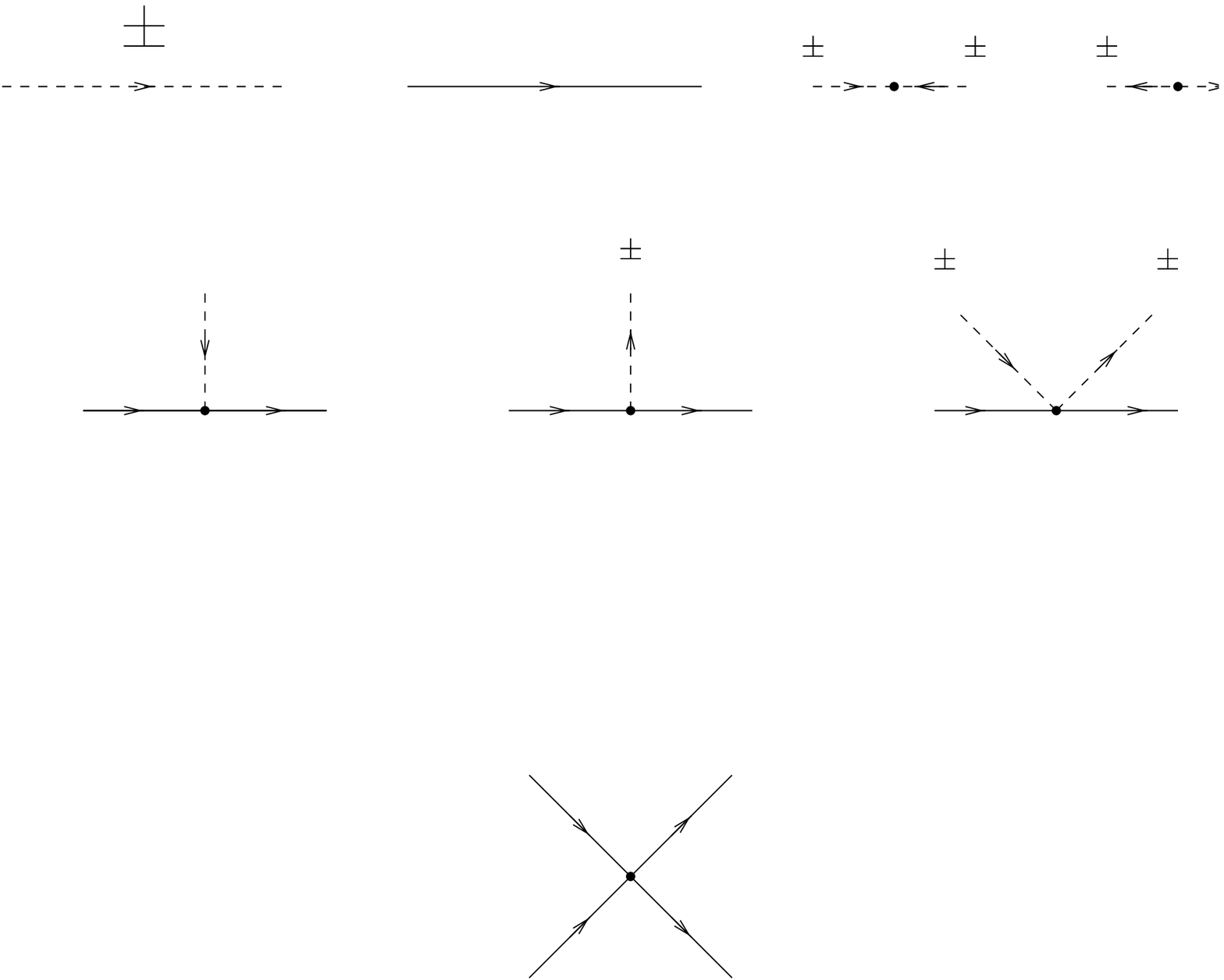}
	\hss}
\begincaption{Figure 3}
Some elements of the Feynman diagrams of the Hamiltonian (2.11)-(2.17).
\endcaption
\endinsert
A dashed line with a plus sign or minus sign represents a magnon 
propagator with ${\cal C}=\pm 1$, whose expression is 
$${i\over q_0-\nu_{\vec q}^\pm+i0^+}\eqno(3.3)$$ for 
${\cal C}=\pm1$. 
A two point vertex with two incoming or two outgoing magnons is 
associated with $\mp 2iJSv_{\vec q}\sin^2{\Theta\over 2}$ for 
${\cal C}=\pm1$. A solid line represents a hole propagator and 
is associated with the expression (2.40). A vertex with an incoming 
and an outgoing holes and an incoming(outgoing) magnon of ${\cal C}=1$
is associated with the factor $i\lambda\sqrt{S\over 2{\cal N}}\tau_-
(\tau_+)$ and that with an incoming(outgoing) magnon of ${\cal C}=-1$ 
with $i\lambda\sqrt{S\over 2{\cal N}}\rho_1\tau_-(\tau_+)$. A vertex 
with two hole lines and two magnon lines of the same ${\cal C}$ is 
associated with a factor $i{\lambda\over 2{\cal N}}\tau_3$. A Coulomb 
vertex with two incoming and two outgoing electrons is associated with 
$-i{2\over {\cal N}}(D_{\vec q}^{(0)}+D_{\vec q}^{(0)\prime}\rho_1)$ 
with $\vec q$ the momentum being transferred in the scattering, where the 
exchange term has been neglected. Other vertices of 
(2.11)-(2.17) may also be read off easily but the above ingredients are 
all we need for the following calculations.

\subsection{3.1 The ground state energy}

The free energy of the ground state per site in the grand canonical 
ensemble comes from the magnetic terms, (2.12) and the Fermi sea and reads
$$f=2S^2(J^\prime+2J\cos\Theta)+{1\over {\cal N}}\sum_{\vec p}\theta
(-\epsilon_{\vec p})\epsilon_{\vec p}.\eqno(3.4)$$ Diagrammatically, the Fermi
sea term corresponds to $i\times$ a closed hole loop. The canting angle is 
determined by the minimization condition
$$\Big({\partial f\over\partial\Theta}\Big)_\mu=0,\eqno(3.5)$$ and we obtain
$$\cos{\Theta\over 2}={\lambda t\over 32JS}\sum_{\vec p}
{v_{\vec p}\over\Delta_{\vec p}}\theta(-\epsilon_{\vec p}).\eqno(3.6)$$ 
Diagrammatically, this condition eliminates the sum of tadpole diagrams
to the order of tree level of the magnon lines and one loop level of 
hole lines. The hole density $n_h$, the number of holes per site, 
is related to the chemical potential 
through $$n_h=\Big({\partial f\over\partial\mu}\Big)_\Theta
={1\over {\cal N}}\sum_{\vec p}\theta(-\epsilon_{\vec p}).
\eqno(3.7)$$ The solution of (3.6) and (3.7) determines the canting angle 
and the chemical potential as a function of the hole density. By a Legendre 
transformation, we obtain the canonical ground state energy
$$u=f+\mu n_h=2(J^\prime+2J\cos\Theta)S^2+{1\over {\cal N}}\sum_{\vec p}
\theta(-\epsilon_{\vec p})(\epsilon_{\vec p}+\mu),\eqno(3.8)$$
in which the chemical potential and the canting angle are expressed in 
terms of $n_h$ with the aid of (3.6) and (3.7). With a strong 
double exchange, $\lambda S>>t, t^\prime$ 
$$\epsilon_{\vec p}=-{1\over 2}\lambda S-t^\prime u_{\vec p}-tv_{\vec p}\cos
{\Theta\over 2}-\mu\eqno(3.9)$$ and the sum over $\vec p$ is 
restricted near the bottom of the band at low hole densities $n_h<<1$,
and an explicit expression of $\Theta$ follows from (3.6), i.e. 
$$\cos{\Theta\over 2}={n_ht\over 4JS^2}.\eqno(3.10)$$ This can be 
recounciled with the formula (1.8) by setting $b=n_ht$. Substituting this 
into (3.8) we find $$u=2(J^\prime-2J)S^2-({1\over 2}
\lambda S+2t^\prime)n_h-{t^2\over 2JS^2}n_h^2.\eqno(3.11)$$
The expression of (3.10) and (3.11) was also obtained by de Gennes [5]
with the magnetic Hamiltonian (1.3).

\subsection {3.2 The magnon spectrum}

Following the standard strategy, we introduce two self energy functions
$\Pi_{\pm}(q_0,\vec q)$ and $\Lambda_{\pm}(q_0,\vec q)$, where
$\Pi_{\pm}(q_0,\vec q)$ represents the sum of all amputated one 
particle irreducible diagrams with an incoming and an outgoing magnons 
of ${\cal C}=\pm1$ and $\Lambda_{\pm}(q_0,\vec q)$ the sum of all one 
particle irreducible diagrams with two incoming (two outgoing) magnons
of ${\cal C}=\pm1$. Then the full propagator with an incoming and an 
outgoing magnons reads 
$${\cal D}_{\pm}(q_0,\vec q)={-iF_{\pm}(-q_0,-\vec q)
\over -F_{\pm}(q_0,\vec q)F_{\pm}(-q_0,-\vec q)+\Lambda_{\pm}^2
(q_0,\vec q)}\eqno(3.12)$$ and that with two incoming (two outgoing) 
magnons reads
$${\cal V}_{\pm}(q_0,\vec q)=-i{\Lambda_{\pm}(q_0,\vec q)
\over -F_{\pm}(q_0,\vec q)F_{\pm}(-q_0,-\vec q)+\Lambda_{\pm}^2
(q_0,\vec q)},\eqno(3.13)$$ where 
$$F_{\pm}(q_0,\vec q)=q_0-\nu_{\vec q}^\pm-\Pi_{\pm}(q_0,\vec q)
.\eqno(3.14)$$ The magnon spectrum is determined by the pole condition:
$$-F_{\pm}(q_0,\vec q)F_{\pm}(-q_0,-\vec q)+\Lambda_{\pm}^2
(q_0,\vec q)=0.\eqno(3.15)$$ The derivation of (3.12) and (3.13) follows 
the same steps as for dressed propagators of an interacting boson 
system with a Bose condensate[9].

The self energy functions have been computed to the order of one 
hole loop and the relevant diagrams are exhibited in Fig. 4. 

\topinsert
\hbox to\hsize{\hss
	\epsfxsize=4.0truein\epsffile{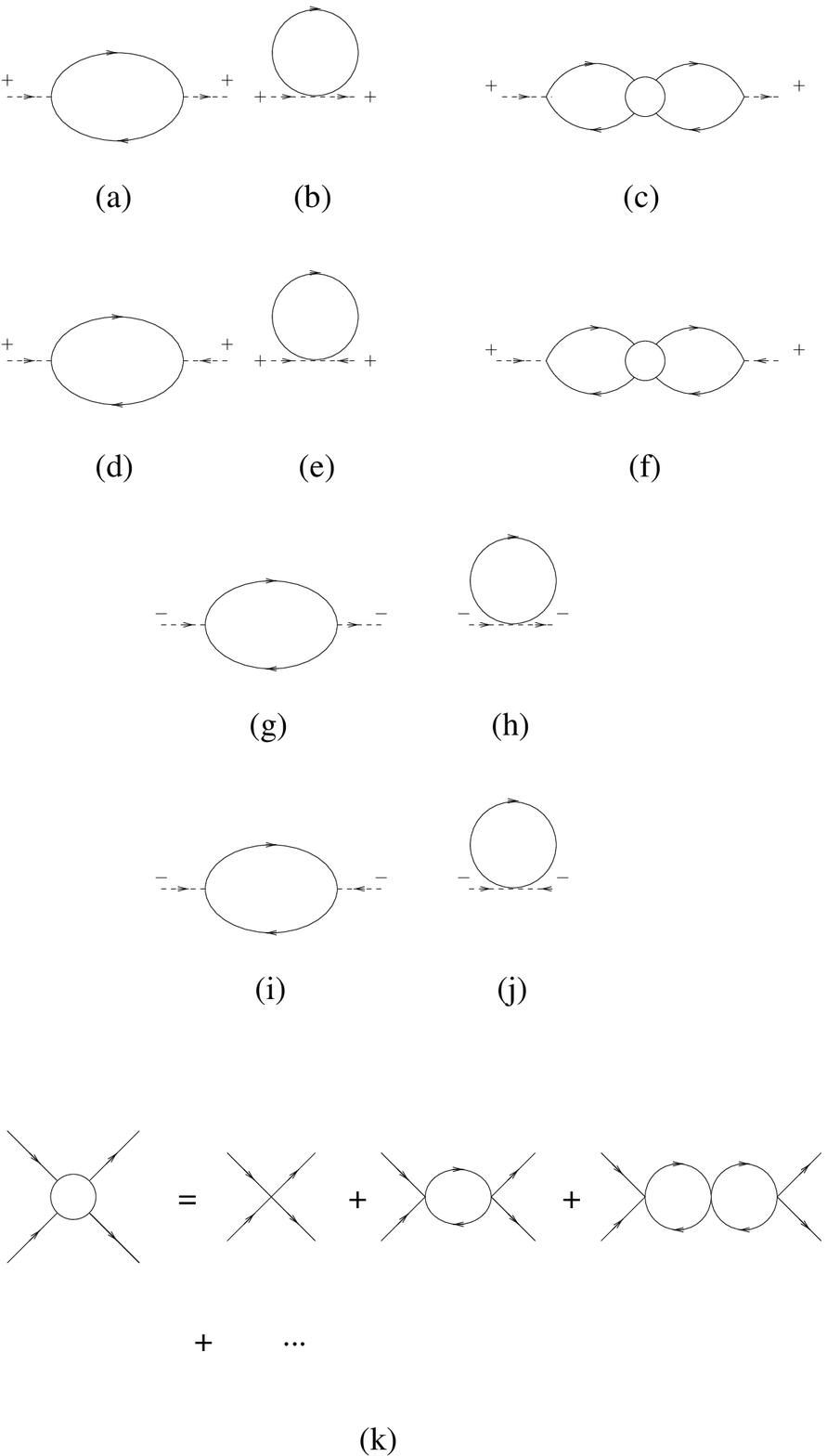}
	\hss}
\begincaption{Figure 4}
Feynmann diagrams for the magnon self energy functions $\Pi_\pm(q_0,\vec q)$ 
and $\Lambda_\pm(q_0,\vec q)$. The trivial diagrams giving rise to the 
first terms of (3.22) and (3.26) are not displayed.
\endcaption
\endinsert

Note that the Debye screening of the long range Coulomb interaction 
has been included in the ${\cal C}=1$ branch since it plays a 
crucial rule to stablize the canting order. The computation 
is straightforward and we list in the following 
the main results in the approximation $\lambda S>>t,t^\prime$

{\it{${\cal C}=1$ Branch}}[10]: 
$$\Pi_{+}(q_0,\vec q)=\Pi_{+}^{(0)}(q_0,\vec q)+\Pi_{+}^{(1)}(q_0,\vec q)
,\eqno(3.16)$$ where $\Pi_+^{(0)}(q_0,\vec q)$ denotes (a)$+$(b), 
the part without Coulomb vertices, and is given by
$$\Pi_+^{(0)}(q_0,\vec q)={\lambda\over 2{\cal N}}\sum_{\vec p}
\theta(-\epsilon_{\vec p})\times$$ $$\times
\Big[1-{\lambda S\over \lambda S+q_0+t^\prime(u_{\vec p}-u_{\vec p-\vec q})
+t(v_{\vec p}+v_{\vec p-\vec q})\cos{\Theta\over 2}}\Big]$$
$$+{t^2\over 2{\cal N}S}\sin^2{\Theta\over 2}\sum_{\vec p}v_{\vec p
-{\vec q\over 2}}^2{\theta(-\epsilon_{\vec p+{\vec q\over 2}})-
\theta(-\epsilon_{\vec p-{\vec q\over 2}})\over 
\epsilon_{\vec p+{\vec q\over 2}}-\epsilon_{\vec p-{\vec q\over 2}}-q_0}
\eqno(3.17)$$ and $\Pi_+^{(1)}(q_0,\vec q)$ comes from 
the screened Coulomb force,
$$\Pi_+^{(1)}(q_0,\vec q)=({\rm{c}})=
{t^2\over 2S}C^2(q_0,\vec q)D(q_0,\vec q)
\sin^2{\Theta\over 2},\eqno(3.18)$$ where $C_(q_0,\vec q)$ corresponds to 
the coupling between a magnon of ${\cal C}=1$ to the density fluctuation 
and reads $$C_(q_0,\vec q)={1\over {\cal N}}\sum_{\vec p}
v_{\vec p-{\vec q\over 2}}{\theta(-\epsilon_{\vec p+{\vec q\over 2}})-
\theta(-\epsilon_{\vec p-{\vec q\over 2}})\over 
\epsilon_{\vec p+{\vec q\over 2}}-\epsilon_{\vec p-{\vec q\over 2}}-q_0}
\eqno(3.19)$$ and ${\cal D}(q_0,\vec q)$ corresponds to the dressed 
Coulomb propagator $$D(q_0,\vec q)=({\rm{k}})=
{e^2D_{\vec q}^{(0)}\over \varepsilon+e^2D_{\vec q}^{(0)}\sigma(q_0,
\vec q)}\eqno(3.20)$$ with the density-density correlation 
function, $\sigma(q_0,\vec q)$, given by $$\sigma(q_0,\vec q)=
-{1\over {\cal N}}\sum_{\vec p}{\theta(-\epsilon_{\vec p+{\vec q\over 2}})-
\theta(-\epsilon_{\vec p-{\vec q\over 2}})\over 
\epsilon_{\vec p+{\vec q\over 2}}-\epsilon_{\vec p-{\vec q\over 2}}-q_0}
.\eqno(3.21)$$ Likewise, we have $$\Lambda_{+}(q_0,\vec q)=2JSv_{\vec q}
\sin^2{\Theta\over 2}+\Lambda_{+}^{(0)}(q_0,\vec q)
+\Lambda_{+}^{(1)}(q_0,\vec q),\eqno(3.22)$$ where $$\Lambda_{+}^{(0)}
(q_0,\vec q)=({\rm{d}})+({\rm{e}})={t^2\over 2{\cal N}S}\sin^2{\Theta\over 2}
\sum_{\vec p}v_{\vec p+{\vec q\over 2}}v_{\vec p-{\vec q\over 2}}\times$$
$$\times{\theta(-\epsilon_{\vec p+{\vec q\over 2}})-
\theta(-\epsilon_{\vec p-{\vec q\over 2}})\over 
\epsilon_{\vec p+{\vec q\over 2}}-\epsilon_{\vec p-{\vec q\over 2}}-q_0}
\eqno(3.23)$$ and $$\Lambda_{+}^{(1)}(q_0,\vec q)=({\rm{f}})={t^2\over 2S}
C(q_0,\vec q)D(q_0,\vec q)C(-q_0,-\vec q)\sin^2{\Theta\over 2}.
\eqno(3.24)$$

{\it{${\cal C}=-1$ Branch}}: 
$$\Pi_-(q_0,\vec q)={\lambda\over 2{\cal N}}\sum_{\vec p}
\theta(-\epsilon_{\vec p})\Big[1-{\lambda S\over 
\lambda S+q_0+t^\prime(u_{\vec p}-u_{\vec p-\vec q})
+t(v_{\vec p}-v_{\vec p-\vec q})\cos{\Theta\over 2}}\Big]$$
$$+{t^2\over 2{\cal N}S}\sin^2{\Theta\over 2}\sum_{\vec p}v_{\vec p
-{\vec q\over 2}}^2\Big[{\theta(-\epsilon_{\vec p+{\vec q\over 2}})
\over -q_0+t^\prime (u_{\vec p-{\vec q\over 2}}-u_{\vec p+{\vec q\over 2}})
-t(v_{\vec p-{\vec q\over 2}}+v_{\vec p+{\vec q\over 2}})
\cos{\Theta\over 2}}$$ $$+{\theta(-\epsilon_{\vec p-{\vec q\over 2}})
\over q_0+t^\prime (u_{\vec p+{\vec q\over 2}}-u_{\vec p-{\vec q\over 2}})
-t(v_{\vec p+{\vec q\over 2}}+v_{\vec p-{\vec q\over 2}})
\cos{\Theta\over 2}}\Big]\eqno(3.25)$$ and $$\Lambda_-(q_0,\vec q)=
-2JSv_{\vec q}\sin^2{\Theta\over 2}-{t^2\over 2{\cal N}S}\sin^2{\Theta\over 2}
\sum_{\vec p}v_{\vec p+{\vec q\over 2}}v_{\vec p-{\vec q\over 2}}\times$$
$$\times\Big[{\theta(-\epsilon_{\vec p+{\vec q\over 2}})
\over -q_0+t^\prime (u_{\vec p-{\vec q\over 2}}-u_{\vec p+{\vec q\over 2}})
-t(v_{\vec p-{\vec q\over 2}}+v_{\vec p+{\vec q\over 2}})
\cos{\Theta\over 2}}$$ $$+{\theta(-\epsilon_{\vec p-{\vec q\over 2}})
\over q_0+t^\prime (u_{\vec p+{\vec q\over 2}}-u_{\vec p-{\vec q\over 2}})
-t(v_{\vec p+{\vec q\over 2}}+v_{\vec p-{\vec q\over 2}})
\cos{\Theta\over 2}}\Big]\eqno(3.26)$$
It follows from the equilibrium condition (3.6) for the canting angle that 
$$F_{\pm}(0,p\hat z)+\Lambda_{\pm}(0,p\hat z)=0.\eqno(3.27)$$ 
The case with $\vec p=0$ of this equation is the consequence 
of the Goldstone theorem and (3.27) is found to hold without the large 
$\lambda S$ approximation. Furthermore we find $$F_-(0,0)-\Lambda_-(0,0)=
0,\eqno(3,28)$$ $$F_+(0,0)-\Lambda_+(0,0)=-16JS\sin^2{\Theta\over 2}$$
$$+{1\over {\cal N}}{t^2\over 2S}\sin^2{\Theta\over 2}
(<v_{\vec p}^2>-<v_{\vec p}>^2)\sum_{\vec p}
\delta(-\epsilon_{\vec p}),\eqno(3.29)$$ where $<...>$ denotes the 
average over the Fermi surface
$$<F_{\vec p}>\equiv{\sum_{\vec p}F_{\vec p}\delta(-\epsilon_{\vec p})
\over\sum_{\vec p}\delta(-\epsilon_{\vec p})}.\eqno(3.30)$$
A simple expression of the magnon spectrum can be obtained at 
low carrier density $n_h<<1$ and with a parabolic approximation of the 
band. The $q_0$ dependence of $\Pi_{\pm}(q_0,\vec q)$ and $\Lambda_{\pm}
(q_0,\vec q)$ may be dropped under the condition that $q_0<<v_Fq$ with 
$v_F$ the Fermi velocity. The second term of (3.29) may also be neglected 
since it is smaller than the first term by a factor of the order of 
$n_h^{2\over 3}$ and eq. (3.27) becomes 
$$q_0^2-[\nu_{\vec q}^\pm+\Pi_\pm(0,\vec q)]^2+\Lambda_\pm^2(0,\vec q)
=0\eqno(3.31)$$ with $q_0$ replaced by $\omega_{\pm}$ we obtain the 
following long wavelength formula for the magnon energy:
$$\omega_+=4\alpha_\perp\sin{\Theta\over 2}\sqrt{JS
\Big[-J^\prime S+{n_h\over 4S}t^\prime+{tt^\prime n_h\sin^2{\Theta\over 2}
\sec{\Theta\over 2}\over 4S(t^\prime+t\cos{\Theta\over 2})}\Big]}
\eqno(3.32)$$ for ${\cal C}=+1$ and
$$\omega_-=\alpha_\perp\sqrt{(-J^\prime S
+{n_h\over 4S}t^\prime\sec^2{\Theta\over 2})
\Big[(-J^\prime S+{n_h\over 4S}t^\prime)\alpha_\perp^2+JS\alpha^2
\sin^2{\Theta\over 2}\Big]}\eqno(3.33)$$ for ${\cal C}=-1$. $\alpha_\perp$ 
and $\alpha$ of (3.32) and (3.33) are related to the cartesian components of 
$\vec q$ through $$\alpha_\perp^2={l^2\over 2}(q_x^2+q_y^2)\eqno(3.34)$$ and 
$$\alpha^2={l^2\over 2}(q_x^2+q_y^2+2q_z^2).\eqno(3.35)$$

The expressions (3.32) and (3.33) are different, though not far, from that 
obtain by de Gennes using the magnetic Hamiltonian (1.3). Both branches 
contribute to the specific heat a power law $T^{3\over 2}$ dependence at 
low $T$. This is different from the observed suppression of the low 
temperature specific heat in the doped samples. It is not likely ,as we 
shall see in the next section that the coupling to a genuine superconducting 
order will improve the situation. An possible explanation is the magnetic 
anisotropicity, which freezes the spin frame to some preferred orientation. 
The detail of this effect will be discussed in the Appendix A.

The approximate equation (3.31) inspires an effective Hamiltonian 
for low momentum magnon spectrum
$$H_{{\rm{mag.}}}=\sum_{\vec q}x_{\vec q}^+\zeta_{\vec q}^\dagger\zeta_{\vec q}
+\mathop{{\sum}'}_{\vec q}y_{\vec q}^+(\zeta_{\vec q}\zeta_{-\vec q}
+\zeta_{-\vec q}^\dagger\zeta_{\vec q}^\dagger)$$ and $$+\sum_{\vec q}
x_{\vec q}^-z_{\vec q}^\dagger z_{\vec q}+\mathop{{\sum}'}_{\vec q}
y_{\vec q}^-(z_{\vec q}z_{-\vec q}+z_{-\vec q}^\dagger z_{\vec q}
^\dagger)\eqno(3.36)$$ with 
$$x_{\vec q}^\pm=\nu_{\vec q}^\pm+\Pi^\pm(0,\vec q)\eqno(3.37)$$
and $$y_{\vec q}^\pm=\Lambda^\pm(0,\vec q).\eqno(3.38)$$ The magnon of 
${\cal C}=-1$ is stable against decay to this order and that of 
${\cal C}=1$ is not. The decay width of the latter is calculated in 
the Appendix C and is found much smaller than the energy (3.32), for 
large $S$ and small $n_h$.

The most remarkable lesson we learned from above Green function 
formalism is that the role played by the long range Coulomb interaction. 
If the $\Pi_+^{(1)}(q_0,\vec q)$ term of (3.16) were not there, 
$\Pi_+(q_0,\vec q)$ would be dominated by the second term of 
$\Pi_+^{(0)}(q_0,\vec q)$, which is of the order $n_h^{{1\over3}}$ and 
is negative. This would make (3.29) positive for ${\cal C}=1$ and the 
corresponding solution for magnon energy imaginary.
This is associated with the negative coefficient of the 
$n_h^2$ term of the bulk energy (3.11), which would be unstable against 
clustering if the holes were electrically neutral. It is the long range 
Coulomb interaction that turns (3.29) negative and a real magnon spectrum 
follows. This effect can not be obtained from the magnetic 
Hamiltonian (1.3).

\section{4. Monte Carlo Simulation of the effective 
	 Magnetic Hamiltonian}

The calculations of the last two sections are restricted to the zero 
temperature. For finite temperatures, numerical simulation is the only 
means of exploring the phase transitions. The Hamiltonian (1.1), 
involving both spin and fermion variables is prohibitively complicated  
to perform a Monte Carlo simulation on a sufficiently large lattice. 
Therefore we switch to the effective magnetic Hamiltonian (1.3) with 
classical spins. The argument which justify such an approximation 
can be found in [6].

The partition function corresponding to the Hamiltonian (1.3) is 
$$Z=\int \prod_jd^2\vec S_je^{-{H_{\rm{eff.}}\over k_BT}}\eqno(4.2)$$ 
where we have made classical approximation of the spins and the 
integration $d^2\vec S_j$ is over the solid angle of $\vec S_j$ with 
its magnitude fixed. The internal energy per magnetic site is 
given by 
$$u=k_BT^2{\partial\over \partial T}\ln Z\eqno(4.3)$$ 
and the specific heat per site is given by
$$c_V={\partial u\over \partial T}.\eqno(4.4)$$

The high temperature expansion of the internal energy per 
site, $u$, obtained by expanding the integrand of (5.2) to a power 
series of ${H_{{\rm{eff.}}}\over k_BT}$ and the first few terms 
give rise to 
$$u=-{4\over 3}(b^\prime+2b)-{2\over 3}{A^\prime+2A\over k_BT}
+{8\over 3}{B\over (k_BT)^2}+O((k_BT)^{-3}),\eqno(4.5)$$
where $$A=J^2S^4-{4\over 5}JbS^2+{1\over 6}b^2,\eqno(4.6)$$ 
$$A^\prime=J^{\prime 2}S^4-{4\over 5}J^\prime b^\prime S^2
+{1\over 6}b^{\prime 2},\eqno(4.7)$$ and 
$$B=(JS^2-{2\over 5}b)^2(J^\prime S^2-{2\over 5}b^\prime)
+({6197\over 48000}-{9\over 64}G)b^2b^\prime\eqno(4.8)$$ with $G$ the 
Catalan's constant given by
$$G=\sum_{n=0}^\infty{(-1)^n\over (2n+1)^2}=0.9160.\eqno(4.9)$$ 
The corresponding specific heat reads
$$c_V=\Big[{2\over 3}{A^\prime+2A\over (k_BT)^2}-{16\over 3}
{B\over (k_BT)^3}+O((k_BT)^{-4})\Big]k_B.\eqno(4.10)$$

At low temperature, equal partition theorem implies that
$$u=u_0+k_BT\eqno(4.11)$$ and the specific heat approaches $k_B$. 
and therefore can not be trusted there.

Extensive Monte carlo simulation has been performed on this 
system. There are four parameters in the magnetic Hamiltonian and 
three dimensionless ratio can be formed: $J^\prime/J\le 1$, which 
measure the anisotropicity of the antiferromagnetic exchange; 
$b^\prime/b$, which measure the anisotropicity of the double exchange, 
and $b/JS^2$, which measure the strength of the double exchange relative 
to the AF exchange. All these ratios are expected to be doping dependent.
It follows from the specific heat data for $x=0$ that $J^\prime\sim J$,  
The specific heat data also indicate that
significant anisotropicity is developed with doping. Therefore, we 
simulated a number of $J^\prime/J$ ratio for $b=b^\prime=0$ and only 
two ratios $J^\prime/J=-0.4,0.5$ are considered for $b, b^\prime\neq 0$.
In the latter case, we are more interested in the interplay between 
the double exchange and AF exchange for various $b$ and $b^\prime$ is 
chosen to equal to $b$ for the sake of simplicity. Most of the results 
except the finite size scaling behavior are extracted from a lattice of 
2000 sites. The heat bath algorithm was used for the simulations 
without double exchange and Metropolis [11] algorithm was used for 
the simulations with double exchange.

	Fig. 5 shows the phase diagram for $J^\prime/J=-0.4, 0.5$, 
obtained from the MC study as compared with the corresponding mean field 
result of de Gennes [5]. The magnitude of the transition temperature for 
various $b$ is reduced from MF by about a factor 2 but the demarkation 
point between AF dominance and FM dominance remains close to the MF 
value $b/JS^2=2.5$. The specific heat extracted from long MC runs 
are plotted in Fig.6 for several $b$ value at $J^\prime/J=-0.4$. 

\topinsert
\hbox to\hsize{\hss
	\epsfxsize=4.0truein\epsffile{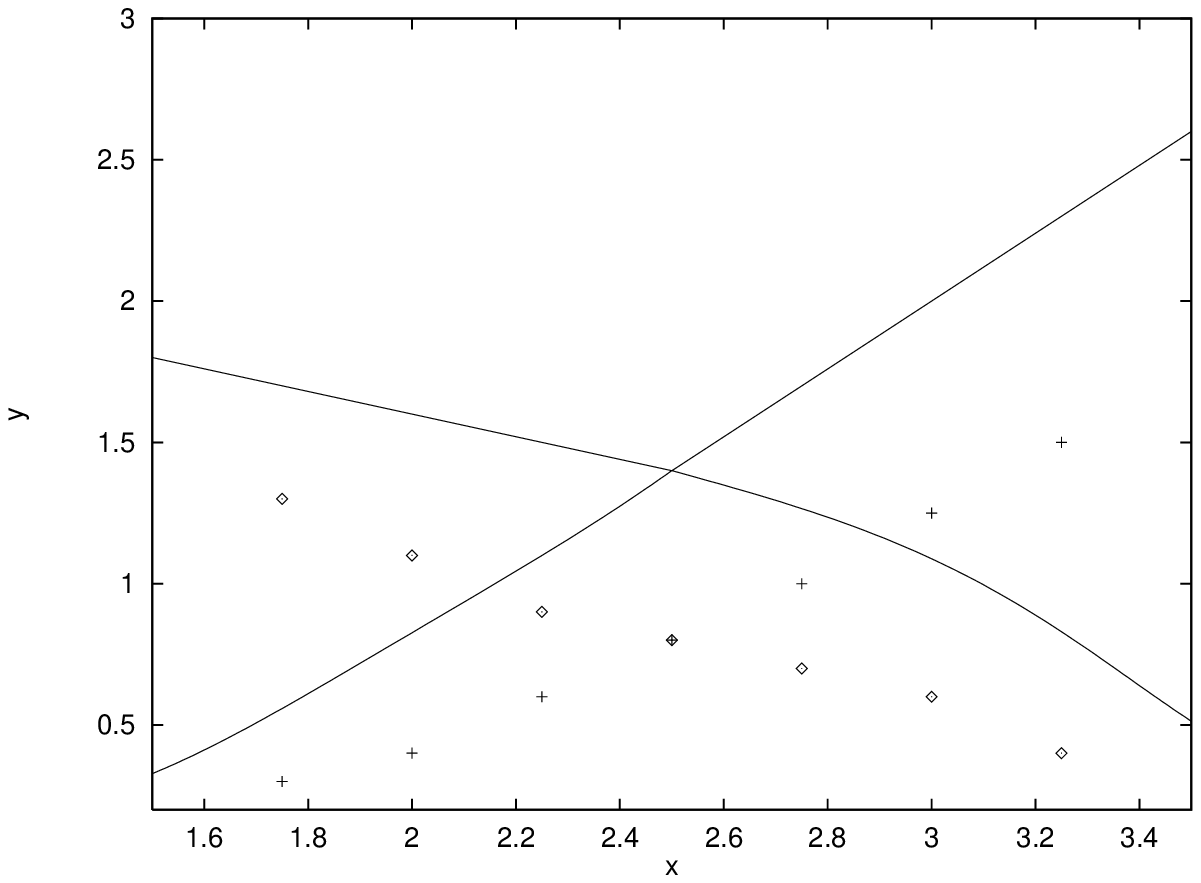}
	\hss}
\centerline{(a)}
\hbox to\hsize{\hss
	\epsfxsize=4.0truein\epsffile{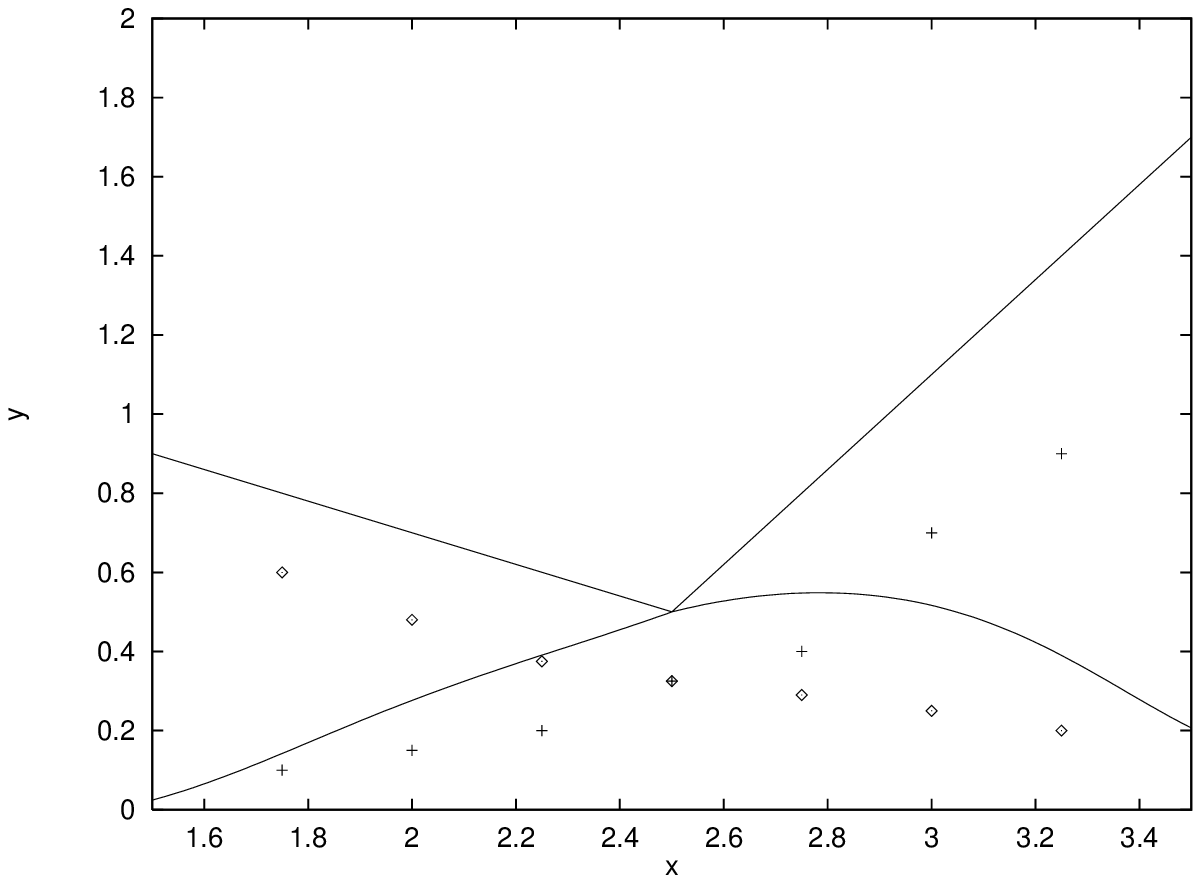}
	\hss}
\centerline{(b)}
\begincaption{Figure 5}
The Monte Carlo results of the magnetic phase diagrams for (a) 
${J^\prime\over J}=-0.4$; (b) ${J^\prime\over J}=0.5$. The solid lines 
is given by the molecular field approximation [5]. The diamonds denote 
the onset of an antiferromagnetic order and the crosses denote the 
onset of a ferromagnetic order. $x={b\over JS^2}$, $y={k_BT\over 5J}$ 
with $k_B$ the Boltzmann constant.
\endcaption
\endinsert

\topinsert
\hbox to\hsize{\hss
	\epsfxsize=4.0truein\epsffile{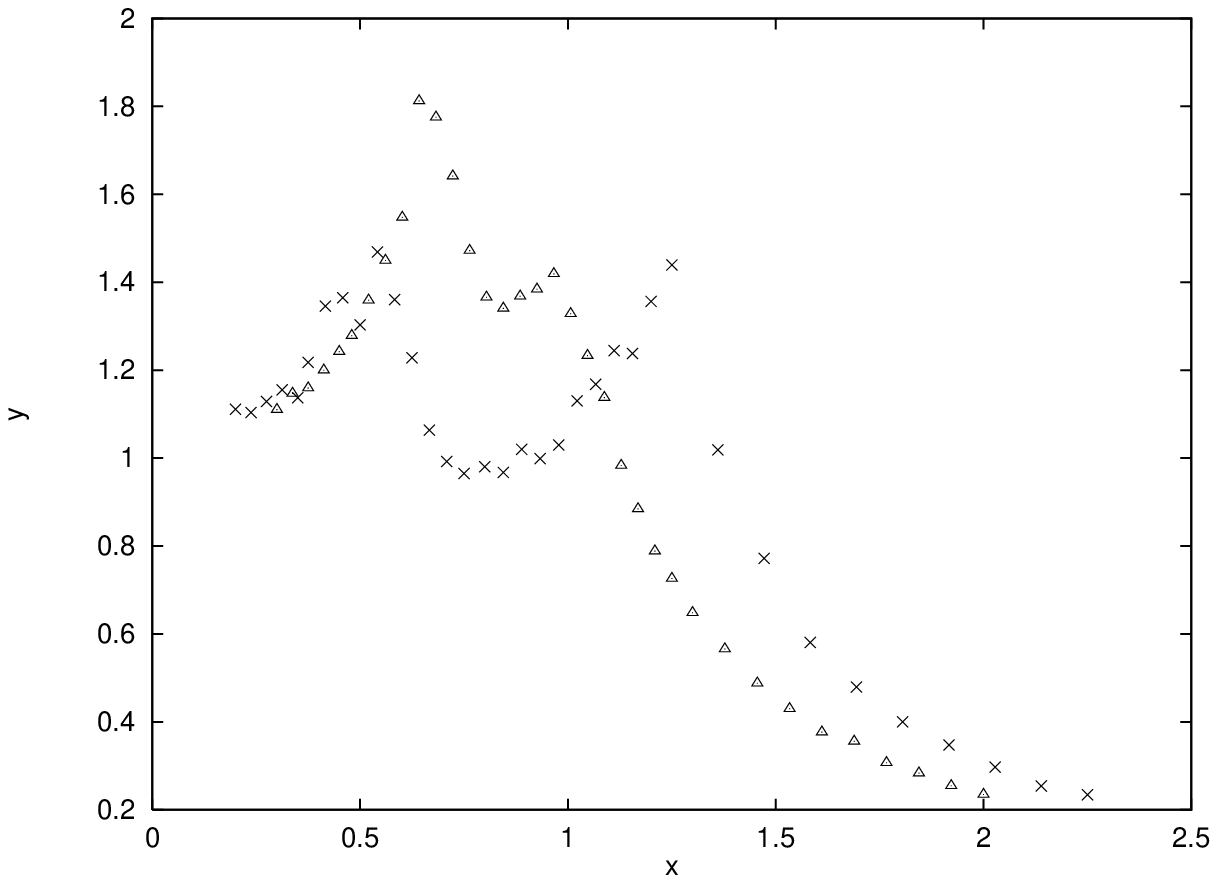}
	\hss}
\centerline{(a)}
\hbox to\hsize{\hss
	\epsfxsize=4.0truein\epsffile{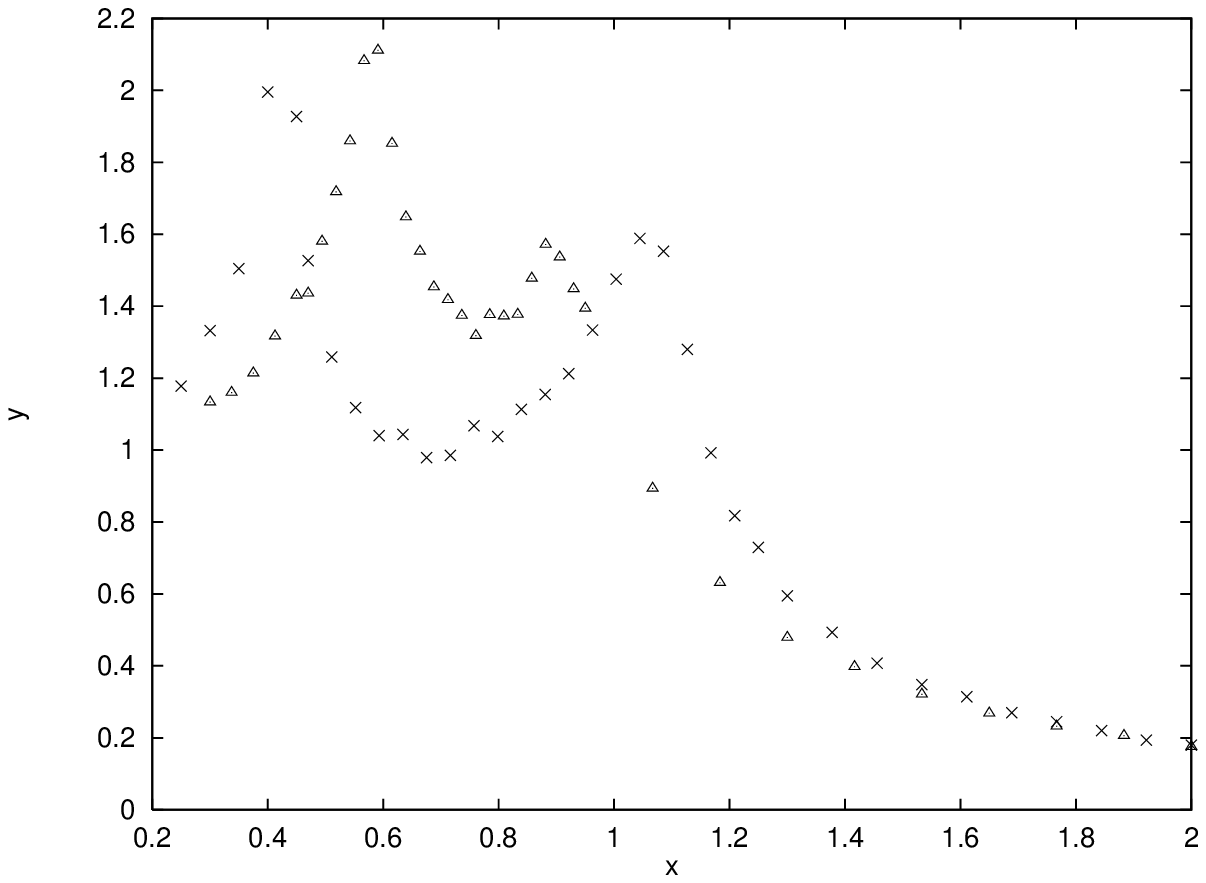}
	\hss}
\centerline{(b)}
\begincaption{Figure 6}
The Monte Carlo results for the specific heat for several strengths of the 
double exchange and ${J^\prime\over J}=-0.4$; $x={k_BT\over 5J}$ and 
$y=c_V$ in arbitrary unit. (a) ${b\over JS^2}=3$
(cross), 2.75(triangle); (b) ${b\over JS^2}=2$(cross), 2.25(triangle).
\endcaption
\endinsert

The two peaks corresponds to the transitions from disordered phase 
to the FM(AF) phase at higher $T$ and the transitions from FM phase 
to the canting phase at lower $T$. 24000 configurations are included 
in the average for the data points. The estimated error ranges from 
one percent outside the critical region to ten percent inside the 
the region. As the two transition approach
each other, its always the peak at higher $T$ that is suppressed. This 
is consistent with the experimental fact that the higher $T$ peak is 
somehow less sharper than the other one[2]. The scaling behavior as the 
lattice size increases from ${\cal N}=250$ to ${\cal N}=6750$ is 
displayed in Fig. 7 and the growth of these peaks indicate two 
genuine phase transition as ${\cal N}\to\infty$. 

\topinsert
\hbox to\hsize{\hss
	\epsfxsize=4.0truein\epsffile{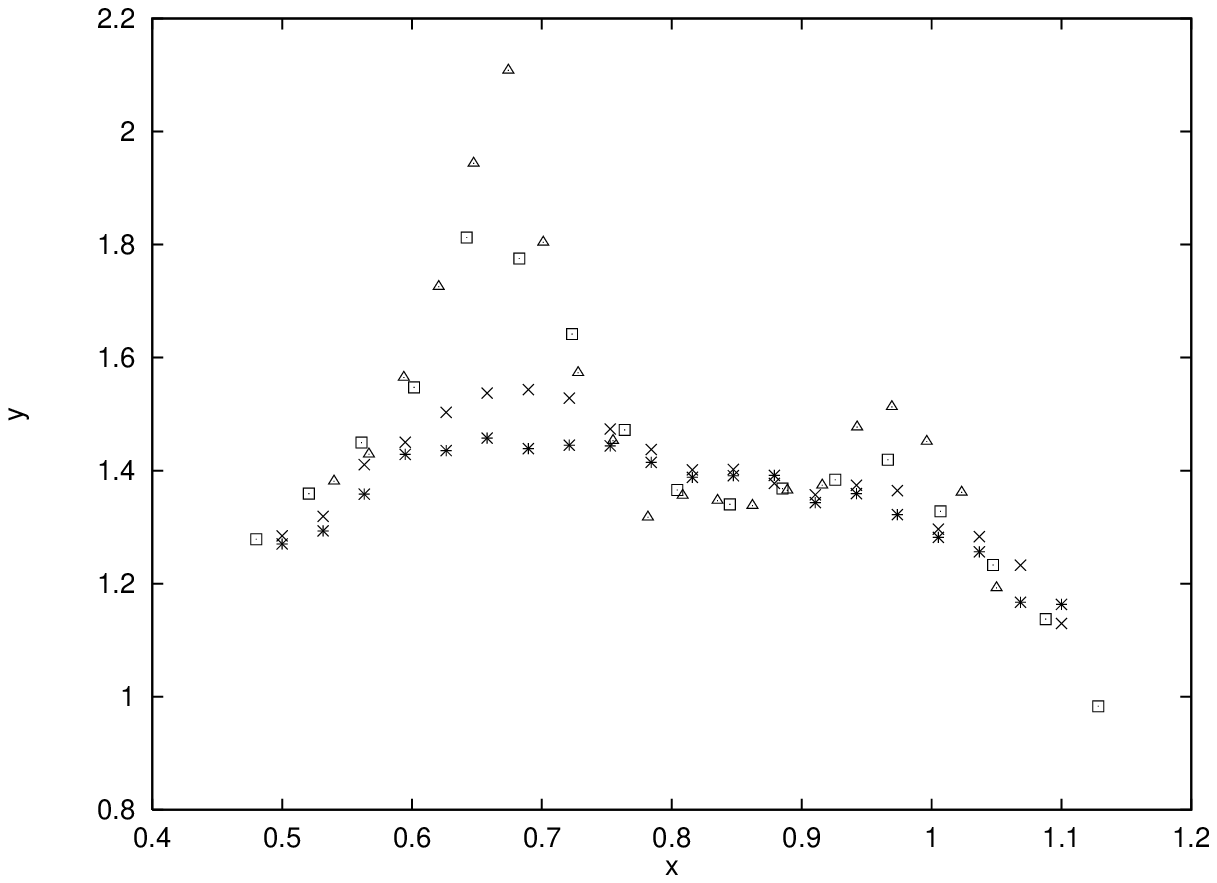}
	\hss}
\begincaption{Figure 7}
The scaling behavior of the specific heat for ${\cal N}=250$(star),
432(cross), 2000(square) and 6750(triangle) near the transitions,
where ${J^\prime\over J}=-0.4$, ${b\over JS^2}=2.75$, $x={b\over JS^2}$
and $y=c_V$ in arbitrary unit.
\endcaption
\endinsert

	The canting order can be viewed as a superposition of FM and 
AF orders. As the temperature is lowered for $b/JS^2<2.5$, the 
system enter this phase from AF order. In another word, a FM order 
is triggered at the lower transition temperature. On the other hand, 
as the temperature is lowered for $b/JS^2>2.5$, the system enter 
the canting phase from FM order and a AF order is triggered at the 
lower transition temperature. Since a homogeneous magnetic field couple 
directly to the FM order only, we expect that the peak at lower $T$ is 
sensitive to the external magnetic field for $b/JS^2<2.5$ and the 
peak at higher $T$ is sensitive for $b/JS^2>2.5$. This is indeed 
the case for the MC results of the specific heat with an external 
magnetic field shown in Fig.8. 

\topinsert
\hbox to\hsize{\hss
	\epsfxsize=4.0truein\epsffile{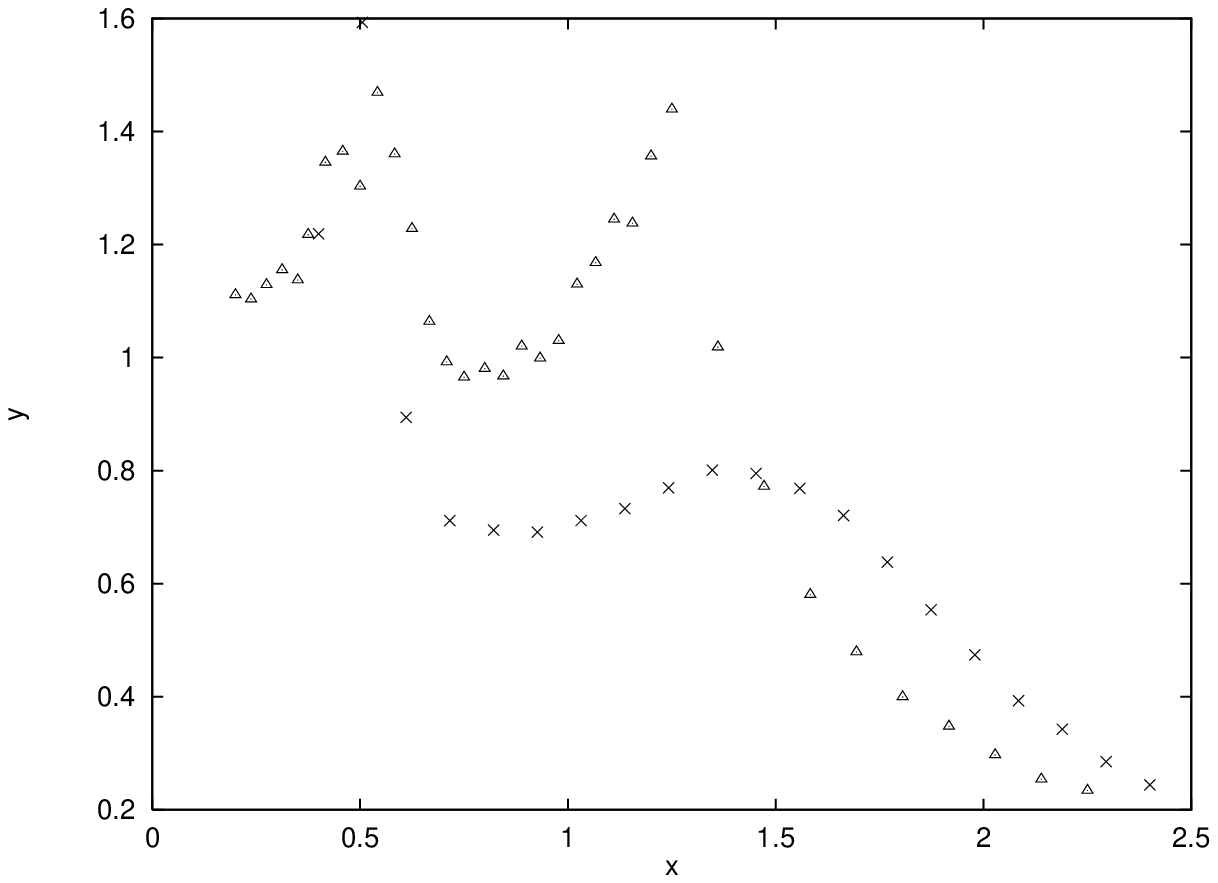}
	\hss}
\centerline{(a)}
\hbox to\hsize{\hss
	\epsfxsize=4.0truein\epsffile{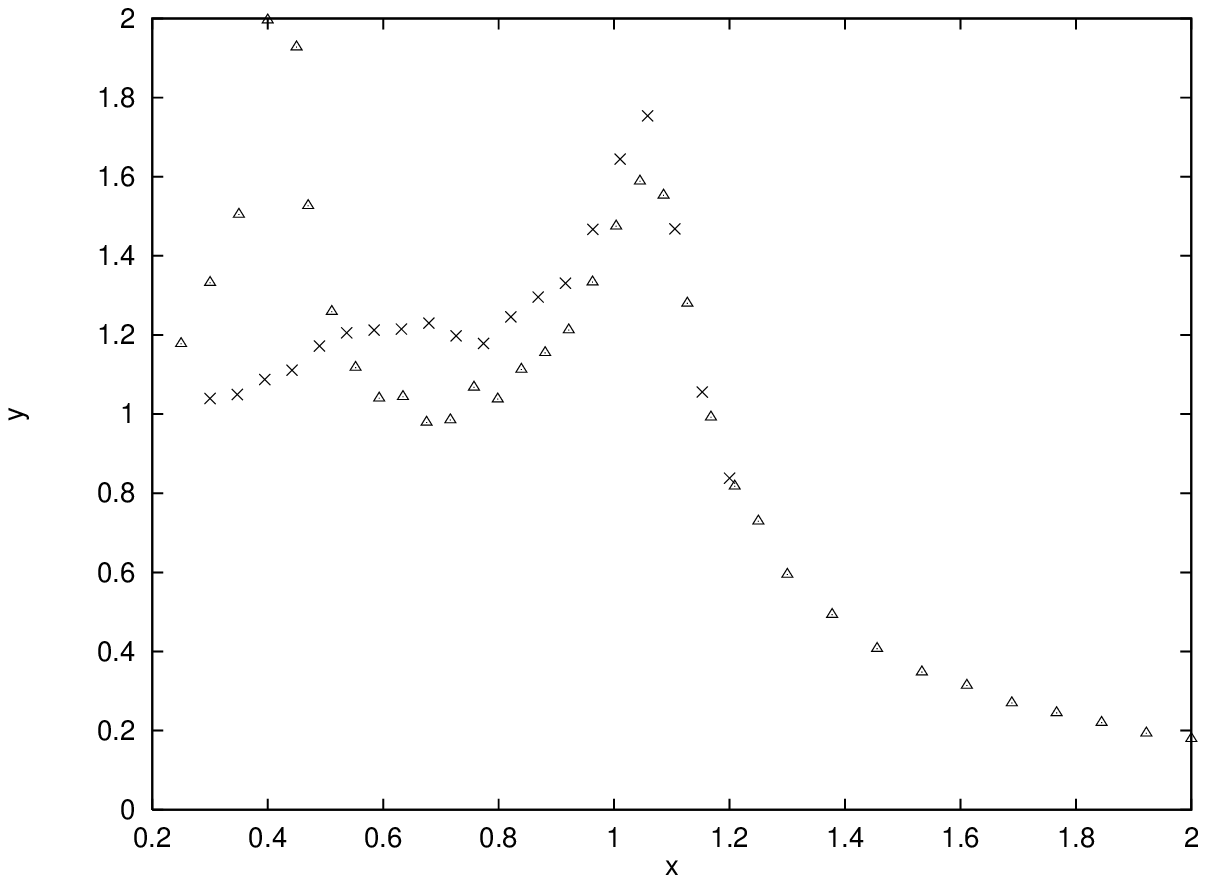}
	\hss}
\centerline{(b)}
\begincaption{Figure 8}
The effect of an external magnetic field on the specific heat peaks for 
${J^\prime\over J}=-0.4$ and ${b\over JS^2}=3(a)$, 2(b). The triangles 
correspond to zero external field, ${\mu B\over JS^S}=0$ and the crosses
to ${\mu B\over JS^2}=0.5$; $x={k_BT\over 5J}$ and $y=c_V$ in arbitrary 
unit.
\endcaption
\endinsert

\section{5. Coupling to the Superconducting Order}

Under the assumption that only the band with ${\cal C}=1$ and energy
(3.1), referred to as the conducting band, is partially filled, the 
superconducting pairing, if there is any, has to be of odd parity. 
The spin orientation of the two partners of a 
pairing with momenta $\vec p$ and -$\vec p$ are parallel, like the 
pairing in the $A$ phase of $He^3$[12]. But, the orientation itself depends 
on the magnitude of $\vec p$. The collective mode (Bogoliubov mode) 
originated from the long range order has the ${\cal C}$ number one and 
is liable to couple with the magnon of the same channel. 

Introduce the annihilation and creation operators of the itinerant holes, 
$a_{\vec ps}^\iota$ and $a_{\vec p}^{\iota\dagger}$ with $\iota$ and $s$ 
ladling the quantum numbers of the four bands ($\iota=\pm$ for ${\cal C}
=\pm1)$, the hole-magnon coupling term of the Hamiltoninan can be written as
$$H_3=-\sqrt{{1\over {\cal N}}}\sum_{\vec p,\vec p^\prime,\iota,s}[\lambda_
{\vec p,\vec q,s}^\iota\zeta_{\vec q}a_{\vec p^\prime s}^{\iota\dagger}
a_{\vec ps}^\iota+\lambda_{\vec p,\vec q,s}^{\iota*}\zeta_{\vec q}^\dagger 
a_{\vec ps}^{\iota\dagger} a_{\vec p^\prime s}^\iota+...],\eqno(5.1)$$
where we show explicitly only the terms preserving the ${\cal C}$ number of 
the holes. $$\lambda_{\vec p,\vec p^\prime s}^\iota=\sqrt{S\over 2}\lambda 
\phi_{\vec p^\prime s}^{\iota\dagger}\sigma_-\phi_{\vec ps}^\iota\eqno(5.2)$$ 
with $\phi_{\vec ps}^\iota$ the $2\times 1$ wave function defined in (2.36).
Consider the scatering process of two holes with opposite momenta
$$h(\vec p)+h(-\vec p)\to h(\vec p^\prime)+h(-\vec p^\prime)$$ with 
$\vec q=\vec p^\prime-\vec p$ and $|\vec q|<<|\vec p|$. The second order 
perturbation in $H_3$ produce an energy dependent effective potential 
mediating the two holes, i.e.
$$U_{{\rm{eff.}}}=-i\Big[\lambda_{\vec p^\prime,\vec p}
\lambda_{-\vec p^\prime,-\vec p}^*[{\cal D}(q_0,\vec q)+{\cal V}
(q_0,\vec q)]$$ $$-\lambda_{-\vec p^\prime,\vec p}\lambda_{\vec p^\prime,
-\vec p}^*[{\cal D}(P_0,\vec P)+{\cal V}(P_0,\vec P)]\eqno(5.3)$$
with $P_0=p_0+p_0^\prime$ and $\vec P=\vec p+\vec p^\prime$, which consists 
of the direct interaction (the first term) and the exchange interaction 
(the second term). For $\vec p$ and $\vec p^\prime$ near the 
Fermi surface and $\vec q\to 0$, the direct term dominate and 
we have $$U_{{\rm{eff.}}}={8t^2\over S}\sin^2{\Theta\over 2}{q_0+{
\omega_{\vec q}^2\over 16JS}\csc{\Theta\over 2}\over q_0^2-
\omega_{\vec q}^2},\eqno(5.4)$$ where $\omega_{\vec q}$ is given by 
(3.32) and the superscript has been suppressed and we have also 
substitute the approximation 
$$\lambda_{\vec p^\prime,\vec p,s}^\iota=\iota s2\sqrt{{2\over S}}t\sin
{\Theta\over 2}.\eqno(5.5)$$ at low carrier density. The effective 
potential (5.4) is rather similar to the effective potential mediated 
by an acoustic phonon, which is retarded and attractive, and is felt 
by holes in all bands. 
The large core spin $S$ here plays a similar role as the large ion mass 
for the electron(hole)-phonon coupling and we expect that the Migdal theorem 
[13] holds for sufficiently large $S$. Furthermore such an attraction would 
not be there for just ferromagnetic order alone, $\Theta=0$, a point stressed 
also in [14]. There is a simple physical reason: 
Like the electron-phonon coupling, the effective attraction is achieved 
through local polarization of the ions by charge carriers, the attraction 
here is caused by local distortion of the ion spins. The counter 
reaction of this distortion would flip the carriers spin, which 
would cost huge amount of energy $\sim\lambda S$ with only ferromagnetic order, 
since the spin orientation in the conduction band point to a unique 
direction. With canting order, the spin orientation in the conduction 
band varies with the Bloch momentum and the local ion spins can be 
distorted at much less cost of energy. In addition to (5.4), there may be 
other attractions induced by large hole-phonon coupling caused by Jahn-Taylor 
effect. On the other hand, we are still far from inferring that there 
are sufficient attractions to beat the Coulomb repulsion. Furthermore 
the exchange term in (5.4) has to be considered for single specie pairing. 

Leaving aside the pairing mechanism, we may still study the 
coupling of the magnon with the superconducting order by
by adding a pairing term to the Haniltonian (2.11). The total Hamiltonian 
reads now, $${\cal H}=H+\sum_{\vec q}\beta_{\vec q}^\dagger\beta_{\vec q},
\eqno(5.6)$$ where $a_{\vec p}$, $a_{\vec p}^\dagger$ stands for the 
annihilation and creation operators of the itinerant holes, and 
$\beta_{\vec q}$, $\beta_{\vec q}^\dagger$ stands for composite 
boson operators which represent Cooper pairs, i.e., 
$$\beta_{\vec q}={1\over\sqrt{{\cal N}}}\mathop{{\sum}'}_{\vec p} 
g_{\vec p,\vec q}a_{\vec p+{\vec q\over 2}}a_{-\vec p+{\vec q\over 2}}
\eqno(5.7)$$ with $\mathop{{\sum}'}_{\vec p}$ extends over half of the 
Brillouin zone only. The pairing wave function, $g_{\vec p,\vec q}$, 
is odd under the space inversion, i.e.
$$g_{\vec p,\vec q}=-g_{-\vec p,\vec q}\eqno(5.8)$$ on account of the 
anticommutation relation among $a_{\vec p}$'s [15]. The superconductivity 
is implemented through the condensation of the pairing operator of 
zero total momentum, i.e.
$$<S|\beta_{\vec q=0}|S>=\sqrt{{\cal N}}B,\eqno(5.9)$$ where $B$ is the 
long range order parameter and is chosen to be real and positive 
for the sake of simplicity. At the presence of (5.9), the part of the 
total Hamiltonian (5.6), which is quadratic in $a$ and $a^\dagger$ reads
$$H_{SC}=\sum_{\vec p}\epsilon_{\vec p}a_{\vec p}^\dagger a_{\vec p}
+\mathop{{\sum}'}_{\vec p}(\delta_{\vec p}^*a_{\vec p}a_{-\vec p}
+\delta_{\vec p}a_{-\vec p}^\dagger a_{\vec p}^\dagger)\eqno(5.10)$$
with $\epsilon_{\vec p}$ given by (3.1) and 
$$\delta_{\vec p}=Bg_{\vec p,\vec q}|_{\vec q=0}=|\delta_{\vec p}|
e^{i\gamma_{\vec p}}.\eqno(5.11)$$ 
$H_{SC}$ can be diagonalized by the Bogoliubov transformation
$$a_{\vec p}=\alpha_{\vec p}\cos\theta_{\vec p}-e^{i\gamma_{\vec p}}
\alpha_{-\vec p}^\dagger\sin\theta_{\vec p}$$ 
$$a_{-\vec p}^\dagger=e^{-i\gamma_{\vec p}}\alpha_{\vec p}\sin\theta_{\vec p}
+\alpha_{-\vec p}^\dagger\cos\theta_{\vec p},\eqno(5.12)$$ where
$$\cos2\theta_{\vec p}={\epsilon_{\vec p}\over {\cal E}_{\vec p}}$$ 
$$\sin2\theta_{\vec p}={|\delta_{\vec p}|\over {\cal E}_{\vec p}}\eqno(5.13)$$ 
with ${\cal E}_{\vec p}=\sqrt{\epsilon_{\vec p}^2+|\delta_{\vec p}|^2}$. 
and we have $$H_{SC}={1\over 2}\sum_{\vec p}(\epsilon_{\vec p}-
{\cal E}_{\vec p})+\sum_{\vec p}{\cal E}_{\vec p}\alpha_{\vec p}^\dagger
\alpha_{\vec p}. \eqno(5.14)$$ The free energy density of the ground state,
(3.4), is replaced by $$f=2S^2(J^\prime+2J\cos\Theta)+{1\over {\cal N}}
{1\over 2}\sum_{\vec p}(\epsilon_{\vec p}-{\cal E}_{\vec p}).\eqno(5.15)$$
The canting angle is determined now by $$\cos{\Theta\over 2}=
{\lambda t\over 32JS}\sum_{\vec p}{v_{\vec p}\over\Delta_{\vec p}}
{{\cal E}_{\vec p}-\epsilon_{\vec p}\over 2{\cal E}_{\vec p}},\eqno(5.16)$$
and the expression of the carrier density (3.7) becomes $$n_h=
={1\over {\cal N}}\sum_{\vec p}{{\cal E}_{\vec p}-\epsilon_{\vec p}\over 
2{\cal E}_{\vec p}}.\eqno(5.17)$$
In addition to the strong double exchange and low carrier density, we 
assume that the gap energy comparable or smaller than the Fermi energy.
It follows then that the expression (3.10) for $\Theta$ remains an approximate 
solution of (5.16) and (5.17). The reason is that the canting is induced by 
the delocalization of the charge carriers, which lower the energy of the 
kinetic energy by an amount of the order hopping amplitude,
while the approximation amounts to assume that other energy 
scales (Fermi energy and gap energy) are much smaller than that.

The superconducting order affect the magnon spectrum in two ways. 
The first of them amounts to dress the hole lines in Fig. 4
with the long range order. On writing the ${\cal C}=1$ branch of the hole 
propagator, $S^+(p_0,\vec p)$ of (2.40) as 
$$S^+(p_0,\vec p)=i\Big({Z_{\vec p}^+\over p_0-\epsilon_{\vec p}}
+{Z_{\vec p}^-\over p_0-\epsilon_{\vec p}+2\Delta_{\vec p}}\Big),
\eqno(5.18)$$ where $Z_{\vec p}^\pm$ are $2\times 2$ matrices
$$Z_{\vec p}^\pm=\phi_{\vec p\pm}^+\phi_{\vec p\pm}^{+\dagger}=
{1\over 2}\Big(1\mp{tv_{\vec p}\sigma_3^\prime
+{1\over 2}\lambda S\sigma_3\over \Delta_{\vec p}}\Big).\eqno(5.19)$$ 
Only the first term is subject to the modification. The factor 
$${i\over p_0-\epsilon_{\vec p}}=\int_{-\infty}^{\infty}dte^{ip_0t}
<|Ta_{\vec p}(t)a_{\vec p}(0)^\dagger|>\eqno(5.20)$$ of (5.18) ought 
to be replaced by $$\int_{-\infty}^{\infty}dte^{ip_0t}
<S|Ta_{\vec p}(t)a_{\vec p}(0)^\dagger |S>={i(p_0+\epsilon_{\vec p})
\over p_0^2-{\cal E}_{\vec p}^2},\eqno(5.21)$$ where $|S>$ denotes 
the ground state with the long range order, (5.9), and the time evolution 
of $a_{\vec p}(t)$ is governed by the Hamiltonian (5.14). Anomalous 
propagators $$\int_{-\infty}^{\infty}dte^{ip_0t}<S|Ta_{\vec p}(t)
a_{-\vec p}(0)|S>={i\delta_{\vec p}\over p_0^2-{\cal E}_{\vec p}^2} 
\eqno(5.22)$$ and $$\int_{-\infty}^{\infty}dte^{ip_0t}<S|Ta_{-\vec p}(t)
^\dagger a_{-\vec p}(0)^\dagger|S>={i\delta_{\vec p}^*\over p_0^2
-{\cal E}_{\vec p}^2}\eqno(5.23)$$ should also be included. The hole 
propagator of the ${\cal C}=-1$ channel $S^-(p_0,\vec p)$, however, remains 
unchanged. With these modifications, the same magnon spectrum (3.32)
and (3.33) emerge as the leading order of the approximation stated 
following (5.17). 

The second type of influence of the superconducting order is the mixing 
of the magnon of ${\cal C}=1$ and the collective (Bogoliubov mode). 
Instead of an extensive diagrammatic analysis, we take a short cut. 
We first calculate the amplitude of the following two processes:

I: $|i>=\zeta_{\vec q}^\dagger|S>$, $|f>=\beta_{\vec q}^\dagger|S>$;

II: $|i>=\zeta_{\vec q}^\dagger\beta_{-\vec q}^\dagger|S>$, $|f>=|S>$

where, $|i>$ and $|f>$ denote the initial and final states.
Using (4.1), (4.7), we find that 
$$\kappa\equiv\lim_{\vec q\to 0}<f|H_3|i>={tB\over 4{\cal N}\sqrt{S}}
\sin{\Theta\over 2}\sum_{\vec p}|g_{\vec p,0}|^2{{\cal E}_{\vec p}
+\epsilon_{\vec p}\over {\cal E}_{\vec p}^2}.\eqno(5.24)$$

{}From now on, we regard $\beta_{\vec q}$ and $\beta_{\vec q}^\dagger$
as elementary boson operators and introduce the following effective 
Hamiltonian for the mixing
$$H_{{\rm{eff.}}}=\sum_{\vec p}x_{\vec p}\zeta_{\vec p}^\dagger
\zeta_{\vec p}+\mathop{{\sum}'}_{\vec p} y_{\vec p}(\zeta_{\vec p}
\zeta_{-\vec p}+\zeta_{-\vec p}^\dagger\zeta_{\vec p}^\dagger)$$
$$+\sum_{\vec p}\xi_{\vec p}\beta_{\vec p}^\dagger
\beta_{\vec p}+\mathop{{\sum}'}_{\vec p} \eta_{\vec p}(\beta_{\vec p}
\beta_{-\vec p}+\beta_{-\vec p}^\dagger\beta_{\vec p}^\dagger)$$
$$+\kappa\sum_{\vec p}(\beta_{\vec p}^\dagger\zeta_{\vec p}+
\beta_{\vec p}\zeta_{-\vec p}+{\rm{h.c.}}),\eqno(5.25)$$
where $x_{\vec p}$ and $y_{\vec p}$ here are the same as $x_{\vec p}^+$ 
and $y_{\vec p}^+$ in (3.38) and (3.39) and the structure bosonic part 
Hamiltonian can be obtained simply from the boson-fermion model of 
superconductivity [16]. The low momentum limit of the coefficients 
$x_{\vec p}$, $y_{\vec p}$, $\xi_{\vec p}$ and $\eta_{\vec p}$ are all 
positive and satisfy $$x_{\vec p}-y_{\vec p}=s_{\vec p},\eqno(5.26)$$
$$x_{\vec p}+y_{\vec p}=c^2,\eqno(5.27)$$
$$\xi_{\vec p}-\eta_{\vec p}=s_{\vec p}^\prime\eqno(5.28)$$
and $$\xi_{\vec p}+\eta_{\vec p}={\omega_0^2\over \varrho_{\vec p}^\prime}
\eqno(5.29)$$ with $s_{\vec p}$ and $s_{\vec p}^\prime$ two positive quadratic 
form of the momentum $\vec p$. The positivity of $\eta_{\vec q}$ comes 
from the repulsive nature of the boson-boson interaction and the choice 
of the condensate phase made in (5.9). The eqs. (5.26) and (5.28) are the 
consequences of the Goldstone theorem and eq. (5.29) follows from the Coulomb 
interaction with $\omega_0$ the plasma frequency. The set of Heisenberg 
equations of motion read
$$i\dot\zeta_{\vec p}=x_{\vec p}\zeta_{\vec p}+y_{\vec p}\zeta_{-\vec p}
^\dagger+\kappa(\beta_{\vec p}+\beta_{-\vec p}^\dagger),\eqno(5.30)$$
$$-i\dot\zeta_{-\vec p}^\dagger=x_{\vec p}\zeta_{-\vec p}^\dagger+
y_{\vec p}\zeta_{\vec p}+\kappa
(\beta_{\vec p}+\beta_{-\vec p}^\dagger),\eqno(5.31)$$
$$i\dot\beta_{\vec p}=\xi_{\vec p}\beta_{\vec p}+\eta_{\vec p}\beta_{-\vec p}
^\dagger+\kappa(\zeta_{\vec p}+\zeta_{-\vec p}^\dagger),\eqno(5.32)$$
$$-i\dot\beta_{-\vec p}^\dagger=\xi_{\vec p}\beta_{-\vec p}^\dagger
+\eta_{\vec p}\beta_{\vec p}+\kappa
(\zeta_{\vec p}+\zeta_{-\vec p}^\dagger).\eqno(5.33)$$
With the ansatz that $\zeta_{\vec p}\propto e^{-i\Omega t}$,
$\zeta_{-\vec p}\propto e^{i\Omega t}$, $\beta_{\vec p}\propto e^{-i\Omega t}$
and $\beta_{-\vec p}\propto e^{i\Omega t}$, we obtain the secular equation 
for the energy $\Omega$, 
$$\left |\matrix{x_{\vec p}-\Omega&y_{\vec p}&\kappa&\kappa\cr
y_{\vec p}&x_{\vec p}+\Omega&\kappa&\kappa\cr\kappa&\kappa&
\xi_{\vec p}-\Omega&\eta_{\vec p}\cr\kappa&\kappa&\eta_{\vec p}&
\xi_{\vec p}+\Omega\cr}\right |=0\eqno(5.34)$$ and its solutions:
$$\Omega^2={1\over 2}[x_{\vec p}^2-y_{\vec p}^2+\xi_{\vec p}^2-\eta_{\vec p}
^2$$ $$\pm\sqrt{(x_{\vec p}^2-y_{\vec p}^2-\xi_{\vec p}^2+\eta_{\vec p}^2)^2
+4\kappa^2(x_{\vec p}-y_{\vec p})(\xi_{\vec p}-\eta_{\vec p})}].
\eqno(5.35)$$ The stability requires that $E^2>0$ which is equivalent to 
$$(x_{\vec p}+y_{\vec p})(\xi_{\vec p}+\eta_{\vec p})>\kappa^2\eqno(5.36)$$
and is amply guaranteed by the relations (5.27)-(5.29) for $\vec p\to 0$. It 
follows from (5.26)-(5.29) and (5.35) that the low-lying spectrum is given by
$$E^2=c^2\varrho_{\vec p}+O(p^4),\eqno(5.37)$$ where the $O(p^4)$ represents 
the correction from the superconducting order and is therefore rather 
small.

\section{6. Concluding Remarks}

We conclude this paper by a few comments related to the real system, 
$Sr_2YRu_{1-x}Cu_xO_6$ and $Ba_2YRu_{1-x}Cu_xO_6$.

The Monte Carlo simulation and spin wave fitting of the specific data [2] 
for $x=0$ indicates that $J=1.28\sim 2.06$meV. The closeness of the 
locations of the two specific heat suggests that we are in the neighborhood 
of the merging point of the three phases (FM(AF), canting and disorder) 
and implies ${b\over JS^2}=2.5$, which together with (1.8) gives rise to
$t=36.0\sim 57.9$meV. The large double exchange approximation is 
therefore quite good for the real system. Assuming only one component of 
the $d$-orbit of $Ru^{+5}$ gets delocalized, there would be two states 
per site. Making an equal partition of these delocalized states into 
four bands, there would be 0.5 states per site and per band. With only 
one band filled partially, the filling fraction of the band would be 
$4x$. There the low density approximation is not a good one for the 
observed samples (with x=0.07, 0.10). With all three $d$ components 
delocalized and degenerate, the filling fraction can be lowered to 
$4x/3$ and the accuracy of the low density approximation can be 
improved.
                 
So far we have concentrated our attention to the magnetic properties 
of the double exchange system. Now we come to the superconductivity 
of the system. Since it has not been rule out that observed 
superconductivity is caused by the cuprate impurities,
we shall list in the following the pros and the cons to a genuine 
superconductivity of a pure system.

1) Imagine the system $Sr_2YRu_{1-x}Cu_xO_6$ consists of several 
domains of pure phases of the double perovskite 
$Sr_2YRu_{1-x_n}Cu_{x_n}$ and cuprate $YSr_2Cu_{3-y_k}Ru_{y_k}
O_{7-\delta_m}$. Suppose that the size of each domain is sufficient 
large that the interface effect on the stoichiometry may be neglected, 
we expect that $$N=\sum_n N_n+\sum_{k,m}N_{k,m},\eqno(6.1)$$ 
$$N(1-x)=\sum_n N_n(1-x_n)+\sum_{k,m} N_{k,m}y_k,\eqno(6.2)$$
$$Nx=\sum_n N_nx_n+\sum_{k,m}N_{k,m}(3-y_k)\eqno(6.3)$$ and so on, where 
$N$ is the total number of $Y$ ion, $N_n$ is the total number of 
$Sr_2YRu_{1-x_n}Cu_{x_n}O_6$ molecules and $N_{k,m}$ that of 
$YSr_2Cu_{3-y_k}Ru_{y_k}O_{7-\delta_m}$ molecules. The eqs (6.1)-(6.3)
is the consequence of the balance the numbers of $Y$, $Sr$, $Ru$ and $Cu$ 
ions. Combining (6.2) and (6.3) we find $N=\sum_n N_n+3\sum_{k,m}N_{k,m}$.
It follows from (6.1) then that $N_{k,m}=0$, which rules out the 
bulk impurity phase of cuprates.

2) The coherent length of a cuprate superconductor is comparable to 
the lattice spacing. A grain with a small number of unit cells of cuprate is 
already able to sustain a long range order. It may happen that 
a large number of such grains of cuprates get embedded inside the 
bulk double perovskite sample. On the other hand, in order to support 
the bulk resistivity transition, Josephson tunneling between different 
grains has to be significant. This process, however is made rather 
difficult by a canting magnetic order since the splitting of the 
spin degeneracy in the bulk material reduces drastically the 
tunneling probability of a singlet Cooper pair.

3) If the observed superconductivity is a genuine one coexisting 
with the canting magnetic order, the pairing is between the itinerant 
holes with opposite momenta and is of odd parity. The doped $Cu$-ions 
plays the role of impurities, in addition to provide itinerant holes.
(In this aspect, $Sr_2YRu_{1-x}Cu_xO_6$ differs from the cuprates. 
The doping of the latter does not alter the $CuO_2$ planes.)
These impurities offset the degeneracy between opposite momentum and 
thereby suppress the superconductivity. This is similar to the effect 
of magnetic impurities on the singlet pairing. If this factor dominates,
the transition temperature would be lowered with doping, opposite to 
the observation.

Odd parity pairing would be a distinct feature of the superconducting 
order with canting magnetic order and the parity symmetry is 
spontaneously broken, i.e. 
$${\cal P}|S>\neq|S>,\eqno(6.4)$$ where ${\cal P}$ stands for the 
ordinary inversion operator defined by 
$${\cal P}\psi_{\vec p}{\cal P}^{-1}=\psi_{-\vec p},\eqno(6.5)$$
$${\cal P}\zeta_{\vec p}{\cal P}^{-1}=\zeta_{-\vec p},\eqno(6.6)$$
$${\cal P}z_{\vec p}{\cal P}^{-1}=z_{-\vec p}\eqno(6.7)$$
and ${\cal P}^2=1$. Unfortunately, such a parity broken does not 
produce any observational consequence for a homogeneous super phase. 
A new operator which combines the inversion and a charge rotation 
through ${\pi\over 2}$,
$${\cal P}^\prime=e^{i{\pi\over 2}Q}P\eqno(6.6)$$ with $Q$ the charge 
operator, leave the condensate invariant. Acting on the observables 
bilinear in hole operators, momentum, electric current, etc., 
${\cal P}^\prime$ is completely equivalent to ${\cal P}$ since 
these operators carry no charges. The square of ${\cal P}^\prime$, 
however, assumes different eigenvalues for charge even and charge 
odd states, ${\cal P}^{\prime 2}=(-)^Q$. But the states with $Q=$even 
never coher with the states with $Q=$ odd, even with a 
condensate of Cooper pairs of charge $2e$. 

With both an odd parity superconductor and an even parity superconductor
coupled together, none of ${\cal P}$ or ${\cal P}^\prime$ will leave 
the ground state invariant and observational consequences of the odd 
parity pairing can be extract. 

Furthermore, the pairing state $|S>$ breaks also the time reversal 
invariance and this will lead to additional observational 
effects which may serve as an evidence of a genuine 
superconductivity.

Another possible way to distinguish a genuine superconductivity from 
that caused by cuprates is to examine the anisotropicity of the 
observed superconductivity. X-ray data suggested only moderate 
anisotropicity of the crystal structure in contrast to the 
strong anisotropicity of the cuprates because of the quasi-two-
dimensional crystal structure.

In addition to the magnetic properties presented in this paper, 
we have also calculated the usual optical response functions 
on account of the exotic band structure. The motivation came from 
the $\lambda$ term of the Hamiltonian (1.1), which is, with a ferromagnetic 
order canting order, equivalent to an ultra-strong magnetic field 
acting on the spin degrees of freedom. The typical Farady angle 
associated to the optical activity because of $T$-violation of the 
magnetic ordering turns out to be about $10^{-4}$ radians per wavelength 
in the visible region and is characterized 
by a different wavelength dependence from that of the ordinary 
magneto-optical effect. The details of this work will be reported 
in a forthcoming publication [17].

\section{Acknowledgments}

Hai-cang Ren is grateful to Professors T. D. Lee and C. S. Ting 
for discussions. We are indebted to Professors T. K. Lee, 
C. C. Chi and Dr. D. Y. Chen for valuable conversations. We are also 
benefitted from Professor Y. S. Wu, Drs. W. E. Jian and C. C. Wu for their
participation of the early stage of this work. Part of this work 
was carried out during Hai-cang Ren's visit in the National 
Tsing Hua University in 1997 and he would like to thank their 
hospitality. This work is supported in part by U. S. Department 
of Energy under Grant DE-FG02-91ER40651, Task B.

\section{Appendix A}

Within the orthorhombic symmetry, we assume that $\hat z$-axis is the 
easy axis of the ferromagnetic order and $\hat x$-axis is the easy axis 
of the antiferromagnetic order. We add the following anisotropy terms 
to the effective magnon Hamiltonian (3.36)
$$H_{{\rm{mag.}}}^\prime=H_{{\rm{mag.}}}+{w_+\over S}\sum_{<ab>}
[\vec S_{ab+}^2-(\vec S_{ab+}\cdot\hat z)^2]+{w_-\over S}\sum_{<ab>}
[\vec S_{ab-}^2-(\vec S_{ab-}\cdot\hat x)^2],\eqno(A.1)$$ where 
$$\vec S_{ab\pm}={1\over 2}(\vec S_a\pm\vec S_b)\eqno(A.2)$$ and the 
summation extends to all inter-layer bonds. The new terms leaves 
${\cal C}$-parity intact. Applying the large spin expansions (2.9) 
and (2.10) with $\hat\zeta=\hat z$ and $\hat\xi=\hat x$, and making the 
Fourier transformations (B.1) and (B.2) below, we find that
$$H_{mag.}^\prime=\sum_{\vec q}x_{\vec q}^{+\prime}\zeta_{\vec q}^\dagger
\zeta_{\vec q}+\mathop{{\sum}'}_{\vec q}y_{\vec q}^{+\prime}
(\zeta_{\vec q}\zeta_{-\vec q}+\zeta_{-\vec q}^\dagger
\zeta_{\vec q}^\dagger)$$ $$+\sum_{\vec q}x_{\vec q}^{-\prime}
z_{\vec q}^\dagger z_{\vec q}+\mathop{{\sum}'}_{\vec q}
y_{\vec q}^{-\prime}(z_{\vec q}z_{-\vec q}+z_{-\vec q}^\dagger z_{\vec q}
^\dagger),\eqno(A.3)$$ where
$$x_{\vec q}^{\pm\prime}=x_{\vec q}^\pm+{1\over 2}w_+(4\mp v_{\vec q})
\Big(1+\cos^2{\Theta\over 2}\Big)+{1\over 2}w_-(4\pm v_{\vec q})+{1\over 2}
(4\mp v_{\vec q})w_-\sin^2{\Theta\over 2}\eqno(A.4)$$ and 
$$y_{\vec q}^{\pm\prime}=y_{\vec q}^\pm-{1\over 2}w_+(4\mp v_{\vec q})
\sin^2{\Theta\over 2}-{1\over 2}w_-(4\pm v_{\vec q})+{1\over 2}
w_-(4\mp v_{\vec q})\sin^2{\Theta\over 2}\eqno(A.5)$$
with $x_{\vec q}^\pm$ and $y_{\vec q}^\pm$ given in (3.37) and (3.38). 
Goldstone theorem ceases to hold and energy gaps begin to open 
between the ground state and the excitations. The energy gap for the 
magnon of ${\cal C}=1$ reads
$$\delta_+=8\sqrt{2JSw_-}\sin{\Theta\over 2}\eqno(A.6)$$ and that for the 
magnon of ${\cal C}=-1$ reads $$\delta_-=8\sqrt{w_+\Big(w_+\cos^2{\Theta
\over 2}+w_-\sin^2{\Theta\over 2}\Big)}.\eqno(A.7)$$ With a fixed canting 
angle $\Theta$, the spin frame can be viewed as the body frame of a rigid 
rotator and its orientation can be specified by three variables, say 
Euler angles. On the other hand, for small vibrations about the 
laboratory frame (i.e. lattice frame here), the dependence on the 
precession angle is of higher order in magnitude. This is why only 
two parameters $w_\pm$ are included in (A.1).

\section{Appendix B}

\noindent 
{\it{1. Large spin expansion}}

In this part of the appendix, we shall display the steps which lead from 
(1.1) to the expressions (2.11)-(2.17). Making Fourier expansion with respect 
to the lattice Bloch momentum $\vec p$, 
$$c_a=\sqrt{{2\over{\cal N}}}\sum_{\vec p}
c_{\vec p}e^{i\vec p\cdot\vec R_a}\eqno(B.1)$$
$$d_b=\sqrt{{2\over{\cal N}}}\sum_{\vec p}
d_{\vec p}e^{i\vec p\cdot\vec R_b}\eqno(B.2)$$
$$\Psi_a=\sqrt{{2\over{\cal N}}}\sum_{\vec p}
\Psi_{\vec p}^{(a)}e^{i\vec p\cdot\vec R_a}\eqno(B.3)$$
$$\Psi_b=\sqrt{{2\over{\cal N}}}\sum_{\vec p}
\Psi_{\vec p}^{(b)}e^{i\vec p\cdot\vec R_a}\eqno(B.4)$$
with $a$ labeling the sites on $A$ and $b$ labeling the sites on $B$.
Substituting (2.9)-(2.10) and (B.1)-(B.4) into the Hamiltonian (1.1), we 
obtain $$H=H_0+H_1+H_2+H_3+H_4+U_{Coul.},\eqno(B.5)$$
where $$H_0=2{\cal N}S^2(J^\prime+2J\cos\Theta),\eqno(B.6)$$
$$H_1=4JS^{3\over 2}\sqrt{{\cal N}}
(c_0+c_0^\dagger-d_0-d_0^\dagger)\sin\Theta\eqno(B.7)$$
$$H_2=2J^\prime S\sum_{\vec p}(-2+u_{\vec p})
(c_{\vec p}^\dagger c_{\vec p}+d_{\vec p}^\dagger d_{\vec p})$$
$$+2JS\Big[-4\sum_{\vec p}(c_{\vec p}^\dagger c_{\vec p}
+d_{\vec p}^\dagger d_{\vec p})\cos\Theta+\sum_{\vec p}v_{\vec p}
(c_{\vec p}^\dagger d_{\vec p}+d_{\vec p}^\dagger c_{\vec p})
\cos^2{\Theta\over 2}$$ $$-\sum_{\vec p}v_{\vec p}
(c_{\vec p}d_{-\vec p}+d_{-\vec p}^\dagger c_{\vec p}^\dagger)
\sin^2{\Theta\over 2}\Big]+\sum_{\vec p}\Psi_{\vec p}^\dagger
(E_{\vec p}-\mu)\Psi_{\vec p},\eqno(B.8)$$
$$H_3=\lambda\sqrt{{S\over 2{\cal N}}}\sum_{\vec p,\vec q}\Big[
c_{\vec q}\Psi_{\vec p+{\vec q\over 2}}^\dagger\rho_+\vec\tau
\cdot\hat e_A^-\Psi_{\vec p-{\vec q\over 2}}+c_{\vec q}^\dagger
\Psi_{\vec p-{\vec q\over 2}}^\dagger\rho_+\vec\tau
\cdot\hat e_A^+\Psi_{\vec p+{\vec q\over 2}}$$
$$+d_{\vec q}\Psi_{\vec p+{\vec q\over 2}}^\dagger\rho_-\vec\tau
\cdot\hat e_B^-\Psi_{\vec p-{\vec q\over 2}}+d_{\vec q}^\dagger
\Psi_{\vec p-{\vec q\over 2}}^\dagger\rho_-\vec\tau
\cdot\hat e_B^+\Psi_{\vec p+{\vec q\over 2}}\Big]$$
$$+2J\sqrt{{S\over {\cal N}}}\sin\Theta\sum_{\vec p,\vec q}v_{\vec q}
\Big[-c_{\vec q}d_{\vec p+{\vec q\over 2}}^\dagger d_{\vec p-{\vec q\over 2}}
-c_{\vec q}^\dagger d_{\vec p-{\vec q\over 2}}^\dagger d_{\vec p+{\vec q\over
2}}$$ $$+d_{\vec q}c_{\vec p+{\vec q\over 2}}^\dagger c_{\vec p
-{\vec q\over 2}}+d_{\vec q}^\dagger c_{\vec p-{\vec q\over 2}}^\dagger 
c_{\vec p+{\vec q\over 2}}\Big]\eqno(B.9)$$
$$H_4={2J^\prime\over{\cal N}}\sum_{\vec p,\vec p^\prime,\vec q}u_{\vec q}
\Big(c_{\vec p^\prime-{\vec q\over 2}}^\dagger c_{\vec p+{\vec q\over 2}}
^\dagger c_{\vec p^\prime+{\vec q\over 2}}c_{\vec p-{\vec q\over 2}}
+d_{\vec p^\prime-{\vec q\over 2}}^\dagger d_{\vec p+{\vec q\over 2}}^\dagger
d_{\vec p^\prime+{\vec q\over 2}} d_{\vec p-{\vec q\over 2}}\Big)$$
$$+{4J\over{\cal N}}\cos\Theta\sum_{\vec p,\vec p^\prime,\vec q}v_{\vec q}
c_{\vec p^\prime-{\vec q\over 2}}^\dagger d_{\vec p+{\vec q\over 2}}^\dagger
c_{\vec p^\prime+{\vec q\over 2}}d_{\vec p-{\vec q\over 2}}$$
$$-{\lambda\over 2{\cal N}}\sum_{\vec p,\vec p^\prime,\vec q}
\Big(c_{\vec p^\prime-{\vec q\over 2}}^\dagger c_{\vec p^\prime+
{\vec q\over 2}}\Psi_{\vec p+{\vec q\over 2}}^\dagger\rho_
+\vec \tau\cdot\hat\zeta_A\Psi_{\vec p-{\vec q\over 2}}
+d_{\vec p^\prime-{\vec q\over 2}}^\dagger 
d_{\vec p^\prime+{\vec q\over 2}}\Psi_{\vec p+{\vec q\over 2}}^\dagger\rho_-
\vec \tau\cdot\hat\zeta_B\Psi_{\vec p-{\vec q\over 2}}\Big).\eqno(B.10)$$
$$U_{Coul.}={e^2\over {\cal N}\varepsilon}\sum_{\vec p,\vec p^\prime,\vec q}
\Big[D_{\vec q}^{(0)}\Psi_{\vec p+{\vec q\over 2}}^\dagger
\Psi_{\vec p^\prime-{\vec q\over 2}}^\dagger
\Psi_{\vec p^\prime+{\vec q\over 2}}\Psi_{\vec p-{\vec q\over 2}}$$
$$+D_{\vec q}^{(0)\prime}(\rho_3)_{\alpha\beta}(\rho_3)_{\alpha^\prime
\beta^\prime}\Psi_{\vec p+{\vec q\over 2}\alpha}^\dagger
\Psi_{\vec p^\prime-{\vec q\over 2}\alpha^\prime}^\dagger
\Psi_{\vec p^\prime+{\vec q\over 2}\beta^\prime}
\Psi_{\vec p-{\vec q\over 2}\beta}\Big),\eqno(B.11)$$ 
In the above expansion, we have combined hole operators $\Psi_{\vec p}^{(a)}$
and $\Psi_{\vec p}^{(b)}$, each of two components, into a $4\times 1$ column 
matrix, $$\Psi_{\vec p}=\left(\matrix{\Psi_{\vec p}^{(a)}\cr 
\Psi_{\vec p}^{(b)}\cr}\right).\eqno(B.12)$$
Accordingly, $\tau$'s and $\rho$'s stand for the $4\times 4$ Dirac matrices 
(2.18)-(2.19) with $\vec\tau=\tau_1\hat\xi
+\tau_2\hat\eta+\tau_3\hat\zeta$ and $\rho_{\pm}={1\pm\rho_3\over 2}$.
The $4\times 4$ matrix $E_{\vec p}$ is given by 
$$E_{\vec p}=-t^\prime u_{\vec p}-tv_{\vec p}\rho_1+{1\over 2}
\lambda S(\rho_+\hat\zeta_A\cdot\vec \tau
+\rho_-\hat\zeta_B\cdot\vec \tau)\eqno(B.13)$$
and the unit vectors $\hat e_{A(B)}^{\pm}$ is defined as
$$\hat e_A^{\pm}={1\over\sqrt{2}}(\hat\xi_A\pm i\hat\eta)\eqno(B.14)$$
and $$\hat e_B^{\pm}={1\over\sqrt{2}}(\hat\xi_B\pm i\hat\eta).\eqno(B.15)$$
$u_{\vec p}$ and $v_{\vec p}$ in the above equations are given by (2.25) 
and (2.26) 

In addition to invariance under the tetragonal crystal 
rotation, inversion, the following transformation ${\cal C}$, 
$${\cal C}c_{\vec p}{\cal C}^{-1}=-d_{\vec p},\eqno(B.16)$$
$${\cal C}d_{\vec p}{\cal C}^{-1}=-c_{\vec p}\eqno(B.17)$$
and $${\cal C}\Psi_{\vec p}{\cal C}^{-1}=C\Psi_{\vec p}\eqno(B.18)$$ 
with $C=-i\rho_1e^{-{i\pi\over 2}\tau_3}=-\rho_1\tau_3$ also leaves the 
Hamiltonian (B.5)-(B.11) unchanged. This transformation amounts to interchange 
the sublattices $A$ and $B$ while making a $\pi$ rotation of the ion spins 
about $z$-axis. The time reversal invariance is broken by the magnetic 
ordering. This symmetry can be diagonalized via the following unitary 
transformation:
$$c_{\vec p}={1\over \sqrt{2}}(z_{\vec p}+\zeta_{\vec p}),\eqno(B.19)$$
$$d_{\vec p}={1\over \sqrt{2}}(z_{\vec p}-\zeta_{\vec p}),\eqno(B.20)$$
$$\Psi_{\vec p}=V\psi_{\vec p},\eqno(B.21)$$
where $$V={1\over\sqrt{2}}(\rho_-+\rho_+\tau_3)(\rho_1+\rho_3)
\Big(\cos{\Theta\over 4}-i\tau_2\sin{\Theta\over 4}\Big).\eqno(B.22)$$
In terms these new operators, the transformation laws (2.29)-(2.31) 
follow and a simpler form of the Hamiltonian, (2.11)-(2.17), emerges.

\bigskip

\noindent
{\it{B.2, Coulomb vertex}}

In what follows, we shall calculate the Coulomb vertex. Labeling the 
sites on the sublattice $a$ and those on the sublattice by $b$, we have 
$$U_{Coul.}={e^2\over 8\pi\varepsilon}\Big[U(\sum_a:\varrho_a\varrho_a:
+\sum_b:\varrho_b\varrho_b:)$$ $$+\sum_{a\neq a^\prime}{:\varrho_a
\varrho_{a^\prime}:\over|\vec R_a-\vec R_{a^\prime}|}+\sum_{b\neq b^\prime}
{:\varrho_b\varrho_{b^\prime}:\over|\vec R_b-\vec R_{b^\prime}|}
+2\sum_{ab}{:\varrho_a\varrho_b:
\over|\vec R_a-\vec R_b|}\Big],\eqno(B.23)$$ where 
$$\varrho_j=\Psi_j^\dagger\Psi_j,\eqno(B.24)$$ with $j=a,b$ and :(...): 
denotes the normal ordering. Making the Fourier transformation in 
(B.3)-(B.4), we obtain
$$U_{Coul.}={e^2\over 2{\cal N}\varepsilon}\Big[\sum_{\vec q}(S_{\vec q}
+S_{\vec q}^\prime):\varrho_{-\vec q}\varrho_{\vec q}:+\sum_{\vec q}(
S_{\vec q}-S_{\vec q}^\prime):\varrho_{-\vec q}^\prime
\varrho_{\vec q}^\prime:\Big],\eqno(B.25)$$ 
where $$\varrho_{\vec q}=\sum_{\vec p}\Psi_{\vec p+{\vec q\over 2}}
^\dagger\Psi_{\vec p-{\vec q\over 2}}=\sum_{\vec p}
\psi_{\vec p+{\vec q\over 2}}^\dagger\psi_{\vec p-{\vec q\over 2}}
\eqno(B.26)$$ and $$\varrho_{\vec q}^\prime=\sum_{\vec p}
\Psi_{\vec p+{\vec q\over 2}}^\dagger\rho_3\Psi_{\vec p-{\vec q\over 2}}
=\sum_{\vec p}\psi_{\vec p+{\vec q\over 2}}^\dagger\rho_1
\psi_{\vec p-{\vec q\over 2}}\eqno(B.27)$$ 
with $\Psi_{\vec p}$ and $\psi_{\vec p}$ related by the transformation 
(B.21). $S_{\vec q}$ and $S_{\vec q}^\prime$ denote two Madelung type sums, 
$$S_{\vec q}=U+{1\over 4\pi}\sum_{\vec n\neq 0}
{e^{-i\vec q\cdot\vec R_n}\over R_n}\eqno(B.28)$$ and $$S_{\vec q}^\prime=
{1\over 4\pi}\sum_{\vec n}{e^{-i\vec q\cdot\vec R_n^\prime}
\over R_n^\prime}\eqno(B.29)$$ with $$\vec R_n={l\over 2}[n_1(\hat x+\hat y)
+n_2(-\hat x+\hat y)]+n_3l\hat z\eqno(B.30)$$ and $$\vec R_n^\prime=
{l\over 2}\Big[\Big(n_1+{1\over 2}\Big)(\hat x+\hat y)
+\Big(n_2+{1\over 2}\Big)(-\hat x+\hat y)\Big]
+\Big(n_3+{1\over 2}\Big)l\hat z\eqno(B.31).$$It follows that 
$\varrho_{\vec q}^+$ is ${\cal C}$ even and $\varrho_{\vec q}^-$ is ${\cal C}$ 
odd. The whole Coulomb energy is therefore invariant under ${\cal C}$, 
but both branches of the magnons couple to the Coulomb interaction.
$D_{\vec q}^{(0)}$ and $D_{\vec q}^{(0\prime)}$ in the text is given by 
$$D_{\vec q}^{(0)}=S_{\vec q}+S_{\vec q}^\prime\eqno(B.32)$$ and 
$$D_{\vec q}^{(0)\prime}=S_{\vec q}+S_{\vec q}^\prime\eqno(B.33)$$

The calculations of $D_{\vec q}^{(0)}$ and $D_{\vec q}^{(0)\prime}$ at 
low $\vec q$ are similar and we illustrate the method by $D_{\vec q}^{(0)}$.
We first exponentiate the denominator of the summand of (B.28). 
$${1\over R_n}={\sqrt{2}\over l}\int_0^\infty dtt^{-{1\over 2}}
e^{-\pi t(n_1^2+n_2^2+2n_3^3)}\eqno(B.34)$$ and interchange the order of the 
integration and the summation, i.e.
$$S_{\vec q}=U+{\sqrt{2}\over 4\pi l}\int_0^\infty
dt t^{-{1\over 2}}\Big[\vartheta_3\Big({\alpha_1\over 2}\vert it\Big)
\vartheta_3\Big({\alpha_2\over 2}\vert it\Big)
\vartheta_3\Big({\alpha_3\over 2}\vert 2it\Big)-1\Big],\eqno(B.35)$$ where 
$$\vartheta_3(z|\tau)=1+2\sum_{n=1}^\infty e^{in^2\pi\tau}\cos2nz
\eqno(B.36)$$ is one of Jacobian theta function [18]. Using the formula
$$\vartheta_3(z|\tau)=(-i\tau)^{-{1\over 2}}\exp\Big({z^2\over i\pi
\tau}\Big)\vartheta_3\Big({z\over\tau}\vert -{1\over\tau}\Big),\eqno(B.37)$$
we have $$S_{\vec q}=U+{1\over 4\pi l}\int_0^\infty dt t^{-{1\over 2}}
\Big[-\sqrt{2}$$ $$+t^{-{3\over 2}}e^{-{2\alpha_\perp^2+\alpha_3^2\over 8\pi t}}
\vartheta_3\Big(-i{\alpha_1\over 2t}\vert {i\over t}\Big)
\vartheta_3\Big(-i{\alpha_2\over 2t}\vert {i\over t}\Big)
\vartheta_3\Big(-i{\alpha_3\over 4t}\vert {i\over 2t}\Big)\Big].\eqno(B.38)$$ 
Only the "1" term of each $\vartheta_3$ factor of 
(B.38) is sensitive to the limit $\vec q\to 0$ upon integration. By isolating 
out this term and setting $\vec q=0$ for the rest, we arrive at
$$S_{\vec q}=U+{2\over l(2\alpha_\perp^2+\alpha_3^2)}+C,\eqno(B.39)$$
where $$C={1\over 4\pi l}\int_0^\infty\Big[t^{-2}\Big[\vartheta_3^2\Big(0
|{i\over t}\Big)\vartheta_3\Big(0|{i\over 2t}\Big)-1\Big]-\sqrt{{2\over t}}
\Big]=-{0.27214\over l}.\eqno(B.40)$$ Likewise we obtain that
$$S_{\vec q}^{(0)\prime}={2\over l(2\alpha_\perp^2+\alpha_3^2)}
-{0.092711\over l}.\eqno(B.41)$$ Substituting (B.39) and (B.41) into
(B.32) and (B.33), we end up with (2.27) and (2.28).
Only the magnon of ${\cal C}=1$ couple to the long range tail of 
the Coulomb force.

\section{Appendix C}

In this appendix, we calculate the decay rate of the magnon with 
${\cal C}=1$. Neglecting possible couplings to the superconducting 
order, the effective magnon Hamiltonian for ${\cal C}=1$ 
corresponds to the first two terms of () with 
$$x_{\vec p}=\nu_{\vec p}^++\Pi_+(0,\vec p)\eqno(C.1)$$
and $$y_{\vec p}=\Lambda_+(0,\vec p).\eqno(C.2)$$ A mass-shell 
magnon operator is given by the following Bogoliubov transformation:
$$\gamma_{\vec p}=\zeta_{\vec p}\cosh\chi_{\vec p}
+\zeta_{-\vec p}^\dagger\sinh\chi_{\vec p},$$
$$\gamma_{-\vec p}^\dagger=\zeta_{\vec p}\sinh\chi_{\vec p}
+\zeta_{-\vec p}^\dagger\cosh\chi_{\vec p}\eqno(C.3)$$ 
where $$\cosh\chi_{\vec p}={x_{\vec p}\over \omega_{\vec p}^+}$$,
$$\sinh\chi_{\vec p}={y_{\vec p}\over \omega_{\vec p}^+}$$
with $\omega_{\vec p}^+=\sqrt{x_{\vec p}^2-y_{\vec p}^2}$, whose 
low momentum limit is given explicity in (3.32). The amplitude for a 
magnon with momentum $\vec q$ to decay into a pair of an electron 
and a hole with momenta $\vec p\pm{\vec q\over 2}$ near the Fermi
surface, ${\cal A}_{\vec p,\vec q}$, corresponds to the superposition 
of the two sets of diagrams in Fig. 9, i. e.
$${\cal A}_{\vec p,\vec q}={\cal U}_{\vec p+{\vec q\over 2}}^{+\dagger}
\Big(({\rm{Fig. 9a}})\cosh\chi_{\vec q}-({\rm{Fig. 9b}})\sinh\chi
_{\vec q}\Big){\cal U}_{\vec p-{\vec q\over 2}+}^+.\eqno(C.4)$$
With the aid of the Feynman rules listed in the section 3, we find 
that $${\cal A}_{\vec p,\vec q}=i\lambda\sqrt{S\over{2\cal N}}
\phi_{\vec p+{\vec q\over 2}+}^{+\dagger}\Big[\sigma_-\cosh\chi_{\vec q}
-\sigma_+\sinh\chi_{\vec q}$$ $$+2{D^{(0)}_{\vec q}\over\varepsilon+
D^{(0)}_{\vec q}\sigma(\omega,\vec q)}{t\over\lambda S}\sin{\Theta\over 2}
[C(\omega,\vec q)\cosh\chi_{\vec q}-C(-\omega,-\vec q)\sinh
\chi_{\vec q}]\Big]\phi_{\vec p-{\vec q\over 2}+}^+,\eqno(C.5)$$
where the functions $D^{(0)}_{\vec q}$, $\sigma(\omega,\vec q)$ and 
$C(\omega,\vec q)$ are defined in section 3 with $\omega=\omega_{\vec q}^+$. 

Following the same reasoning employed in section 3, we approximate 
$C(\omega,\vec q)$ and $\sigma(\omega,\vec q)$ by $C(0,\vec q)$ and 
$\sigma(0,\vec q)$ and obtain, after some algebra 
$$|{\cal A}_{\vec p,\vec q}|^2={t^2\over{2\cal N}S}\Big[{1\over 2}
(v_{\vec p+{\vec q\over 2}}-v_{\vec p-{\vec q\over 2}})^2\cosh2\chi_{\vec q}
-(v_{\vec p}-<v>)(v_{\vec p+{\vec q\over 2}}-v_{\vec p-{\vec q\over 2}})$$
$$+(v_{\vec p}-<v>)^2(\cosh\chi_{\vec q}
-\sinh\chi_{\vec q})^2\Big]\sin^2{\Theta\over 2}\eqno(C.6)$$ for 
small $\vec q$, where $<..>$ denotes the same Fermi surface average defined 
in (3.30). The decay width is given by
$$\Gamma_{\vec q}=2\pi\sum_{\vec p}|{\cal A}_{\vec p,\vec q}|^2
\theta(\epsilon_{\vec p+{\vec q\over 2}})
\theta(-\epsilon_{\vec p-{\vec q\over 2}})
\delta(\epsilon_{\vec p+{\vec q\over 2}}-\epsilon_{\vec p-{\vec q\over 2}}
-\omega_{\vec q}^+).\eqno(C.7)$$ With parabolic approximation of 
$u_{\vec p}$ and $v_{\vec p}$ and low carrier density, $n_h<<1$, 
$p^2$ is of the order of $n_h^{2\over 3}$ so is the Fermi energy measured 
from the bottom of the conducting band, $\epsilon_F=\mu+2t^\prime+4t\cos
{\Theta\over 2}$. The order of the first term of (C.6) is $n_h^{2\over 3}$, 
the order of the second term is $n_h$ and that of the third term is 
$n_h^{4\over 3}$. Keeping only the first term of 
$|{\cal A}_{\vec p,\vec q}|^2$
and completing the phase space integral for small $\vec q$, we obtain
$$\Gamma_{\vec q}={2\over\pi S}{JStt^{\prime 2}\varepsilon_F\alpha
\sin^2\beta\cos^2\beta\over (t^\prime+t\cos{\Theta\over 2})^2
(t^\prime\sin^2\beta+t\cos{\Theta\over 2})^{3\over 2}\sqrt{t\cos
{\Theta\over 2}}}\tan{\Theta\over 2}\sin^3{\Theta\over 2},\eqno(C.8)$$ 
where we have parametrized the magnon energy $\omega_{\vec q}^+$ by 
$\omega_{\vec q}^+=c\alpha\sin\beta$ with $\alpha=\sqrt{\alpha_1^2+
\alpha_2^2+\alpha_3^2}$ and $\sin\beta=\alpha_\perp/\alpha$. The contribution 
from the second term in the bracket of (C.6) has been neglected in (C.8) since 
it is of higher power of the hole density, $n_h$, when $n_h<<1$. 
As far as the order of magnitude concerned, we have 
$c\sim JS$, $t^\prime\sim t$, and $JS^2\sim tn$. It follows then that
$${\Gamma_{\vec q}\over \omega_{\vec q}}\sim{1\over S}n_h^{2\over 3}<<1
\eqno(C.9)$$ for large spin and small carrier density.

\topinsert
\hbox to\hsize{\hss
	\epsfxsize=4.0truein\epsffile{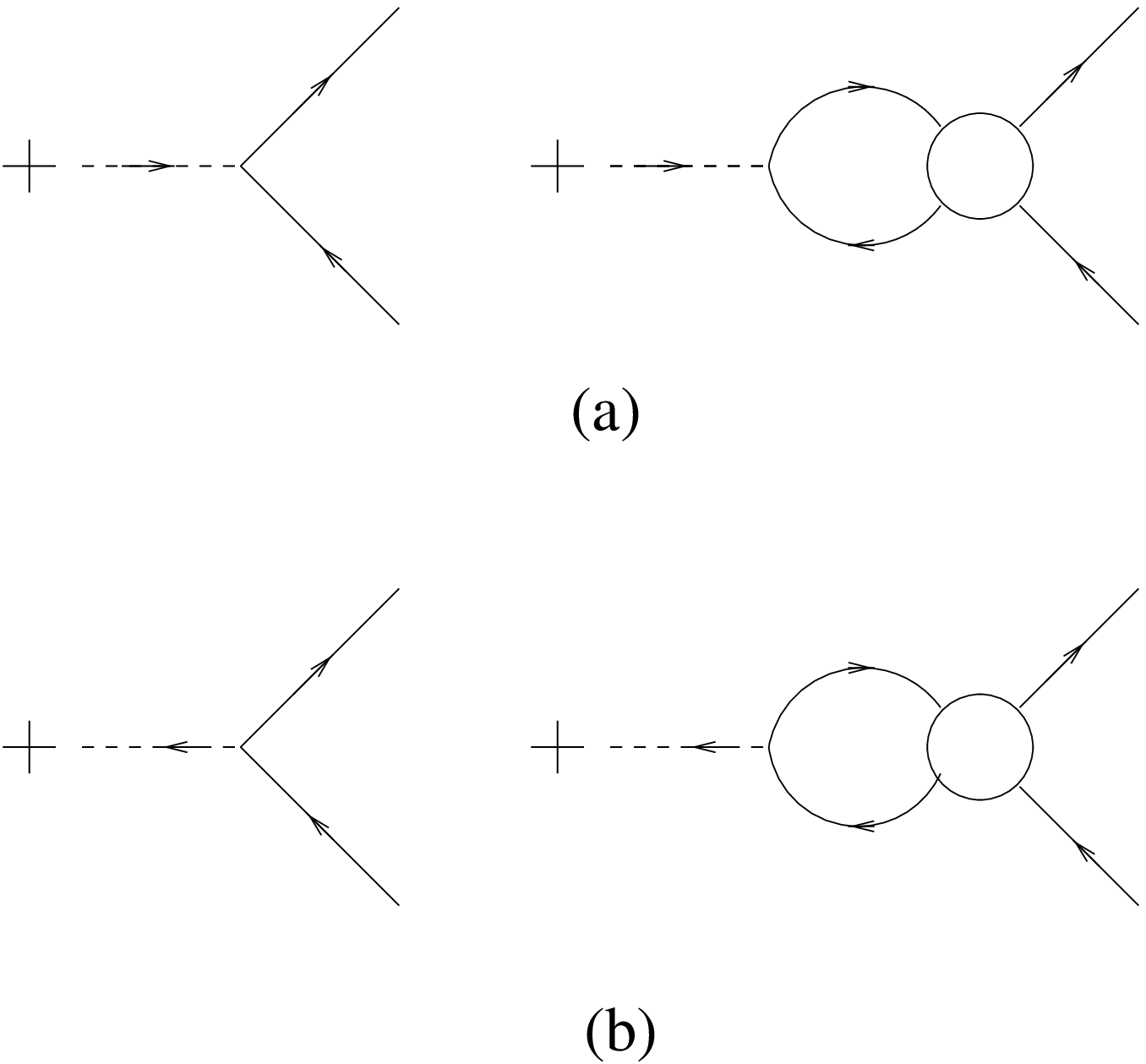}
	\hss}
\begincaption{Figure 9}
The Feynman diagrams for the decay of a magnon with ${\cal C}=1$. The 
open circle with four legs stands for the dressed Coulomb vertex as 
in Fig. 4. 
\endcaption
\endinsert

\references

\ref{1.}{M. K. Wu, et. al., Z. Phys., {\bf 102}, 37 (1997).}
\ref{2.}{M. K. Wu, et. al., submitted to Phys. Rev. Lett..}
\ref{3.}{C. Zener, Phys. Rev., {\bf 82}, 403 (1951).}
\ref{4.}{P. W. Anderson and H. Hasegawa, Phys. Rev., {\bf 100}, 675 (1955).}
\ref{5.}{P. G. de Gennes, Phys. Rev. {\bf 118}, 141 (1960).}
\ref{6.}{A. J. Millis, P. B. Littlewood and B. I. Shraiman, Phys. Rev. 
Lett., {\bf 74}, 5144 (1995).}
\ref{7.}{P. D. Battle and W. I. J. Macklin, J. Solid State Chem., {\bf 52},
138 (1984).}
\ref{8.}{T. Holstein and H. Primakoff, Phys. Rev. {\bf 58}, 1098 (1940).}
\ref{9.}{N. M. Hugenholtz and D. Pines, Phys. Rev. {\bf 116}, 489 (1959).}
\ref{10.}{M. Creutz, 'Quarks, Gluons and Lattices', Cambridge University 
Press, 1983.}
\ref{11.}{To avoid cumbersome expressions, we have suppressed the 
infinitesimal imaginary part of the denominators of $\Pi_+^{(0)}(q_0,
\vec q)$, $\Lambda_+^{(0)}(q_0,\vec q)$, $C(q_0,\vec q)$ and 
$\sigma(q_0,\vec q)$ and supplemented with an equivalent rule that 
the imaginary parts of these functions for real $q_0$ has to be negative,
as is evident from Dyson-Wick perturbation theory.}
\ref{12.}{A. J. Leggett, Rev. Mod. Phys., {\bf 43}, 331 (1975).}
\ref{13.}{See for example, A. L. Fetter and J. D. Walecka, 'Quantum Theory 
of Many Particle System', Chapter 12, (McGrawHill, New York, 1971).}
\ref{14.}{W. E. Pickett, Phys. Rev. Lett., {\bf 77}, 3185 (1996).}
\ref{15.}{Rigorously speaking, pairing term in the Hamiltonian 
(5.6) violates the tetragonal symmetry of the original 
Hamiltonian (1.1). On the other hand, the Goldstone theorem can be 
verified to the leading order of the low carrier density. A rigorous 
treatment of the pairing effect should respect the original symmetry
and therefore should be extended to all bands.}
\ref{16.}{R. Friedberg, T. D. Lee and H. C. Ren, Phys. Rev. {\bf B42}, 
4122 (1990).}
\ref{17.}{H. C. Ren and M. K. Wu, in preparation.}
\ref{18.}{E. T. Whittaker and G. N. Watson, 'A Course of Modern Analysis', 
4th ed., Chapter XXI, Cambridge University Press, 1927.}

\vfill\eject
\end

\bye